\documentclass[aps,pra,groupaddress,twocolumn,showpacs,amsmath]{revtex4}

\usepackage{bm}
\usepackage{graphicx}
\usepackage{latexsym}
\begin{document}

%\title{Quantum trajectory simulation of cavity QED with atomic beams: Atomic\\
%motion and density fluctuations in the weak-excitation limit}
%\title{Cavity QED with thermal atomic beams: Atomic motion\\
%and density fluctuations}
\title{Effect of atomic beam alignment on photon correlation measurements\\ in cavity QED}
\author{L.~Horvath and H.~J.~Carmichael}
\affiliation{Department of Physics, University of Auckland, Private Bag 92019,
Auckland, New Zealand} 
\date{\today}

\begin{abstract}
Quantum trajectory simulations of a cavity QED system comprising an atomic beam traversing
a standing-wave cavity are carried out. The delayed photon coincident rate for forwards scattering
is computed and compared with the measurements of Rempe {\it et al.\/}~[Phys.~Rev.~Lett.~{\bf 67},
1727 (1991)] and Foster {\it et al.\/}~[Phys.~Rev.~A {\bf 61}, 053821 (2000)]. It is shown that
a moderate atomic beam misalignment can account for the degradation of the predicted correlation.
Fits to the experimental data are made in the weak-field limit with a single adjustable
parameter---the atomic beam tilt from perpendicular to the cavity axis. Departures of the
measurement conditions from the weak-field limit are discussed.
\end{abstract}
\pacs{42.50.Pq, 42.50.Lc, 02.70.Uu}

\maketitle

\section{Introduction}
\label{sec:introduction}
Cavity quantum electrodynamics \cite{Berman94,Raimond01,Mabuchi02,Vahala03,Khitrova06,Carmichael07a}
has as its central objective the realization of strong dipole coupling between a discrete transition
in matter (e.g.,~an atom or quantum dot) and a mode of an electromagnetic cavity. Most often strong
coupling is demonstrated through the realization of vacuum Rabi splitting \cite{Mondragon83,Agarwal84}.
First realized for Rydberg atoms in superconducting microwave cavities \cite{Bernardot92,Brune96}
and for transitions at optical wavelengths in high-finesse Fabry Perots \cite{Thompson92,Childs96,Boca04,Maunz05},
vacuum Rabi splitting was recently observed in monolithic structures where the discrete transition
is provided by a semiconductor quantum dot \cite{Yoshie04,Reithmaier04,Peter05},
and in a coupled system of qubit and resonant circuit engineered from superconducting electronics
\cite{Wallraff04}.

More generally, vacuum Rabi spectra can be observed for any pair of coupled harmonic oscillators
\cite{Carmichael94} without the need for strong coupling of the one-atom kind. Prior to observations
for single atoms and quantum dots, similar spectra were observed in many-atom \cite{Raizen89,Zhu90,Gripp96}
and -exciton \cite{Weisbuch92,Khitrova99} systems where the radiative coupling is collectively enhanced.

The definitive signature of single-atom strong coupling is the large effect a single photon in the
cavity has on the reflection, side-scattering, or transmission of another photon. Strong coupling has a
dramatic effect, for example, on the delayed photon coincidence rate in forwards scattering when
a cavity QED system is coherently driven on axis \cite{Carmichael85,Rice88,Carmichael91,Brecha99}.
Photon antibunching is seen at a level proportional to the parameter $2C_1=2g^2/\gamma\kappa$
\cite{Carmichael91}, where $g$ is the atomic dipole coupling constant, $\gamma$ is the atomic spontaneous
emission rate, and $2\kappa$ is the photon loss rate from the cavity; the collective parameter $2C=N2C_1$,
with $N$ the number of atoms, does not enter into the magnitude of the effect when $N\gg1$. In the one-atom
case, and for $2\kappa\gg\gamma$, the size of the effect is raised to $(2C_1)^2$ \cite{Carmichael85,Rice88}
[see Eq.~(\ref{eqn:g2_ideal})].

The first demonstration of photon antibunching was made \cite{Rempe91} for moderately strong coupling
($2C_1\approx4.6$) and $N=18$, $45$, and $110$ (effective) atoms. The measurement has subsequently been
repeated for somewhat higher values of $2C_1$ and slightly fewer atoms \cite{Mielke98,Foster00a}, and
a measurement for one trapped atom \cite{Birnbaum05}, in a slightly altered configuration, has demonstrated
the so-called photon blockade effect \cite{Imamoglu97,Werner99,Rebic99,Rebic02,Kim99,Smolyaninov02}---i.e.,
the antibunching of forwards-scattered photons for coherent driving of a vacuum-Rabi resonance, in which
case a two-state approximation may be made \cite{Tian92}, assuming the coupling is sufficiently strong.

The early experiments of Rempe {\it et al.\/} \cite{Rempe91} and those of Mielke {\it et al.\/}
\cite{Mielke98} and Foster {\it et al.\/} \cite{Foster00a} employ systems designed around a Fabry-Perot
cavity mode traversed by a thermal atomic beam. Their theoretical modeling therefore presents
a significant challenge, since for the numbers of {\it effective\/} atoms used, the atomic beam carries
hundreds of atoms---typically an order of magnitude larger than the effective number \cite{Carmichael99}---into
the interaction volume. The Hilbert space required for exact calculations is enormous ($2^{100}\sim10^{30}$);
it grows and shrinks with the number of atoms, which inevitably fluctuates over time; and the atoms
move through a spatially varying cavity mode, so their coupling strengths are changing in time. Ideally,
all of these features should be taken into account, although certain approximations might be made.

For weak excitation, as in the experiments, the lowest permissible truncation of the Hilbert space---when
calculating two-photon correlations---is at the two-quanta level. Within a two-quanta truncation,
relatively simple formulas can be derived so long as the atomic motion is overlooked \cite{Carmichael91,Brecha99}.
It is even possible to account for the unequal coupling strengths of different atoms, and, through a
Monte-Carlo average, fluctuations in their spatial distribution  \cite{Rempe91}. A significant discrepancy between
theory and experiment nevertheless remains: Rempe {\it et al.\/} \cite{Rempe91} describe how the amplitude
of the Rabi oscillation (magnitude of the antibunching effect) was scaled down by a factor of 4 and a slight
shift of the theoretical curve was made in order to bring their data into agreement with this model;
the discrepancy persists in the experiments of Foster {\it et al.\/} \cite{Foster00a}, except that
the required adjustment is by a scale factor closer to 2 than to 4.

Attempts to account for these discrepancies have been made but are unconvincing. Martini and Schenzle
\cite{Martini01} report good agreement with one of the data sets from Ref.~\cite{Rempe91}; they
numerically solve a many-atom master equation, but under the unreasonable assumption of stationary
atoms and equal coupling strengths. The unlikely agreement results from using parameters that
are very far from those of the experiment---most importantly, the dipole coupling constant is smaller
by a factor of approximately 3.

Foster {\it et al.\/} \cite{Foster00a} report a rather good theoretical fit to one of their 
data sets. It is obtained by using the mentioned approximations and adding a detuning in the calculation
to account for the Doppler broadening of a misaligned atomic beam. They state that ``Imperfect alignment
$\ldots$ can lead to a tilt from perpendicular of as much as $1^\circ$''. They suggest that the mean
Doppler shift is offset in the experiment by adjusting the driving laser frequency and account for
the distribution about the mean in the model. There does appear to be a difficulty with this procedure,
however, since while such an offset should work for a ring cavity, it is unlikely to do so in the presence
of the counter-propagating fields of a Fabry-Perot. Indeed, we are able to successfully simulate
the procedure only for the ring-cavity case (Sec.~\ref{sec:detuning}).

The likely candidates to explain the disagreement between theory and experiment have always
been evident. For example, Rempe {\it et al.\/} \cite{Rempe91} state:

\begin{itemize}
\item[]{
``Apparently the transient nature of the atomic motion through the cavity mode (which is not included
here or in Ref.~[7]) has a profound effect in decorrelating the otherwise coherent response of the sample
to the escape of a photon.''}
\end{itemize}

\noindent
and also:

\begin{itemize}
\item[]{
``Empirically, we also know that $|g^{(2)}(0)-1|$ is reduced somewhat because the weak-field
limit is not strictly satisfied in our measurements.''}
\end{itemize}

\noindent
To these two observations we should add---picking up on the comment in \cite{Foster00a}---that in a
standing-wave cavity an atomic beam misalignment would make the decorrelation from atomic motion a great
deal worse.

Thus, the required improvements in the modeling are: (i) a serious accounting for atomic motion
in a thermal atomic beam, allowing for up to a few hundred interacting atoms and a velocity component
along the cavity axis, and (ii) extension of the Hilbert space to include 3, 4, etc.\ quanta of excitation,
thus extending the model beyond the weak-field limit. The first requirement is entirely achievable in a quantum
trajectory simulation \cite{Carmichael93,Dalibard92,Dum92,Gardiner04,Carmichael07b}, while the second,
even with recent improvements in computing power, remains a formidable challenge.

In this paper we offer an explanation of the discrepancies between theory and experiment in the
measurements of Refs.~\cite{Rempe91} and \cite{Foster00a}. We perform {\it ab initio\/} quantum
trajectory simulations in parallel with a Monte-Carlo simulation of a tilted atomic beam. The parameters
used are listed in Table \ref{tab:parameters}: Set 1 corresponds to the data displayed in Fig.~4(a) of
Ref.~\cite{Rempe91}, and Set 2 to the data displayed in Fig.~4 of Ref.~\cite{Foster00a}. All parameters
are measured quantities--- or are inferred from measured quantities---and the atomic beam tilt alone is varied
to optimize the data fit. Excellent agreement is demonstrated for atomic beam misalignments of
approximately $10\mkern2mu{\rm mrad}$ (a little over $1/2^\circ$). These simulations are
performed using a two-quanta truncation of the Hilbert space.

Simulations based upon a three-quanta truncation are also carried out, which, although not adequate
for the experimental conditions, can begin to address physics beyond the weak-field limit.
From these, an inconsistency with the intracavity photon number reported by Foster {\it et al.\/}
\cite{Foster00a} is found.

\begin{table}[t]
\begin{tabular}{|c||c|c|}
\hline
  Parameter & Set~$1$ & Set~$2$\\
\hline
\hline
\vbox{\vskip3pt\hbox{cavity halfwidth}\vskip3pt\hbox{$\mkern45mu\kappa/2\pi$}\vskip1pt} & \vbox{\hbox{$0.9\mkern1mu{\rm MHz}$}\vskip6pt} & \vbox{\hbox{$7.9\mkern1mu{\rm MHz}$}\vskip6pt}\\
\hline
\vbox{\vskip3pt\hbox{dipole coupling constant}\vskip3pt\hbox{$\mkern70mug_{\rm max}/\kappa$}\vskip1pt} & \vbox{\hbox{$3.56$}\vskip6pt} & \vbox{\hbox{$1.47$}\vskip6pt}\\
\hline 
\vbox{\vskip3pt\hbox{atomic linewidth}\vskip3pt\hbox{$\mkern50mu\gamma/\kappa$}\vskip1pt} & \vbox{\hbox{$5.56$}\vskip6pt} & \vbox{\hbox{$0.77$}\vskip6pt}\\
\hline
\vbox{\vskip3pt\hbox{mode waist}\vskip4pt\hbox{$\mkern33muw_{\rm 0}$}\vskip1pt} & \vbox{\hbox{$50\mkern1mu\mu{\rm m}$}\vskip3pt} & \vbox{\hbox{$21.5\mkern1mu\mu{\rm m}$}\vskip3pt}\\
\hline
\vbox{\vskip3pt\hbox{wavelength}\vskip3pt\hbox{$\mkern35mu\lambda$}\vskip1pt} & \vbox{\hbox{$\mkern5mu852{\rm nm}$ (Cs)}\vskip3pt} & \vbox{\hbox{$\mkern5mu780{\rm nm}$ (Rb)}\vskip3pt}\\
\hline
\hline
\vbox{\vskip3pt\hbox{effective atom number}\vskip4pt\hbox{$\mkern75mu\bar N_{\rm eff}$}\vskip1pt} & \vbox{\hbox{18}\vskip6pt} & \vbox{\hbox{13}\vskip6pt}\\
\hline
\vbox{\vskip3pt\hbox{oven temperature}\vskip3pt\hbox{$\mkern60muT$}\vskip1pt} & \vbox{\hbox{$473\mkern1mu{\rm K}$}\vskip6pt} & \vbox{\hbox{$430\mkern1mu{\rm K}$}\vskip6pt}\\
\hline
\vbox{\vskip3pt\hbox{mean speed in oven}\vskip3pt\hbox{$\mkern65mu\overline{v}_{\rm oven}$}\vskip1pt} & \vbox{\hbox{$ 274.5\mkern1mu{\rm m\!/s}$}\vskip4pt} & \vbox{\hbox{$326.4\mkern1mu
{\rm m\!/s}$}\vskip4pt}\\
\hline
\vbox{\vskip3pt\hbox{mean speed in beam}\vskip3pt\hbox{$\mkern65mu\overline{v}_{\rm beam}$}\vskip1pt} & \vbox{\hbox{$323.4\mkern1mu{\rm m\!/s}$}\vskip4pt} & \vbox{\hbox{$384.5\mkern1mu
{\rm m\!/s}$}\vskip4pt}\\
\hline
\end{tabular}
\caption{ Parameters used in the simulations. Set 1 is taken from Ref.~\cite{Rempe91} and Set 2  from
Ref.~\cite{Foster00a}.}
\label{tab:parameters}
\end{table}                         

Our model is described in Sec.~\ref{sec:cavityQED_atomic_beams}, where we formulate the stochastic master
equation used to describe the atomic beam, its quantum trajectory unraveling, and the two-quanta truncation
of the Hilbert space. The previous modeling on the basis of a stationary-atom approximation is reviewed in
Sect.~\ref{sec:stationary_atoms} and compared with the data of Rempe {\it et al.\/} \cite{Rempe91}
and Foster {\it et al.\/} \cite{Foster00a}. The effects of atomic beam misalignment are discussed in
Sec.~\ref{sec:atomic_beam}; here the results of simulations with a two-quanta truncation are presented. 
Results obtained with a three-quanta truncation are presented in Sec.~\ref{sec:photon_number}, where the
issue of intracavity photon number is discussed. Our conclusions are stated in Sec.~\ref{sec:conclusions}.

\section{Cavity QED with Atomic Beams}
\label{sec:cavityQED_atomic_beams}
\subsection{Stochastic Master Equation: Atomic Beam Simulation}
\label{sec:beam_simulation}
Thermal atomic beams have been used extensively for experiments in cavity QED \cite{Bernardot92,Brune96,Thompson92,Childs96,Raizen89,Zhu90,Gripp96,Rempe91,Mielke98,Foster00a}. The experimental
setups under consideration are described in detail in Refs.~\cite{Brecha90} and \cite{Foster99}. As typically, the beam
is formed from an atomic vapor created inside an oven, from which atoms escape through a collimated opening.
We work from the standard theory of an effusive source from a thin-walled oriface \cite{Ramsey56}, for which for
an effective number $\bar N_{\rm eff}$ of intracavity atoms  \cite{Thompson92,Carmichael99} and cavity mode
waist $\omega_0$ ($\bar N_{\rm eff}$ is the average number of atoms within a cylinder of radius $w_0/2$), the
average escape rate is 
\begin{equation}
R=64\bar N_{\rm eff}\bar v_{\rm beam}/3\pi^2w_0,
\end{equation}
with mean speed in the beam
\begin{equation}
\bar v_{\rm beam}=\sqrt{9\pi k_BT/8M},
\end{equation}
where $k_B$ is Boltzmann's constant, $T$ is the oven temperature, and $M$ is the mass of an atom; the beam
has atomic density
\begin{equation}
\varrho=4\bar N_{\rm eff}/\pi w_0^2l,
\label{eqn:density}
\end{equation}
where $l$ is the beam width, and distribution of atomic speeds
\begin{equation}
P(v)dv=2u^3(v)e^{-u^2(v)}du(v),
\label{eqn:speed_dist}
\end{equation}
$u(v)\equiv 2v/\sqrt\pi\mkern2mu\bar v_{\rm oven}$, where
\begin{equation}
\bar v_{\rm oven}=\sqrt{8k_BT/\pi M}=(8/3\pi)\bar v_{\rm beam}
\end{equation}
is the mean speed of an atom inside the oven, as calculated from the Maxwell-Boltzmann distribution.
Note that $\bar v_{\rm beam}$ is larger than $\bar v_{\rm oven}$ because those atoms that move
faster inside the oven have a higher probability of escape. 

In an open-sided cavity, neither the interaction volume nor the number of interacting atoms is
well-defined; the cavity mode function and atomic density are the well-defined quantities. Clearly, though,
as the atomic dipole coupling strength decreases with the distance of the atom from the cavity axis, those
atoms located far away from the axis may be neglected, introducing, in effect, a finite interaction
volume. How far from the cavity axis, however, is far enough? One possible criterion is to require that
the interaction volume taken be large enough to give an accurate result for the collective coupling strength,
or, considering its dependence on atomic locations (at fixed average density), the {\it probability distribution\/}
over collective coupling strengths. According to this criterion, the actual number of interacting atoms is
typically an order of magnitude larger than $\bar N_{\rm eff}$ \cite{Carmichael99}. If, for example, one
introduces a cut-off parameter $F<1$, and defines the interaction volume by \cite{Carmichael99,Carmichael96,Sanders97}
\begin{equation}
V_F\equiv\{(x,y,z):g(x,y,z)\ge F g_{\rm max}\},
\label{eqn:interaction_volume}
\end{equation} 
with
\begin{equation}
\label{eqn:coupling_strength}
g(x,y,z)=g_{\rm max}\cos(kz)\exp\!\left[-(x^2+y^2)/w_0^2\right]
\end{equation}
the spatially varying coupling constant for a standing-wave TEM$_{00}$ cavity mode \cite{note_cavity_mode}---wavelength
$\lambda=2\pi/k$---the computed collective coupling constant is \cite{Carmichael99}
\begin{equation}
\sqrt{\bar N_{\rm eff}}\mkern5mug_{\rm max}\to\sqrt{\bar N_{\rm eff}^F}\mkern5mug_{\rm max},\nonumber
\end{equation}
with 
\begin{equation}
\bar N_{\rm eff}^F=(2\bar N_{\rm eff}/\pi)\mkern-5mu\left[(1-2F^2)\cos^{-1}F+F\sqrt{1-F^2}\right].
\label{eqn:effective_atom_number}
\end{equation}
For the choice $F=0.1$, one obtains $\bar N_{\rm eff}^F=0.98\bar N_{\rm eff}$, a reduction of the collective
coupling strength by 1\%, and the interaction volume---radius $r\approx3(w_0/2)$---contains approximately
$9\bar N_{\rm eff}$ atoms on average. This is the choice made for the simulations with a three-quanta truncation
reported in Sec.~\ref{sec:photon_number}. When adopting a two-quanta truncation, with its smaller Hilbert space
for a given number of atoms, we choose $F=0.01$, which yields $\bar N_{\rm eff}^F=0.9998\bar N_{\rm eff}$ and
$r\approx4.3(w_0/2)$, and approximately $18\bar N_{\rm eff}$ atoms in the interaction volume on average.

In fact, the volume used in practice is a little larger than $V_F$. In the course of a Monte-Carlo
simulation of the atomic beam, atoms are created randomly at rate $R$ on the plane $x=-w_0\sqrt{|\ln F|}$.
At the time, $t_0^j$, of its creation, each atom is assigned a random position and velocity ($j$ labels a
particular atom),
\begin{equation}
{\bm r}_j(t_0^j)=\mkern-3mu\left(\begin{matrix}-w_0\sqrt{|\ln F|}\\\noalign{\vskip2pt}y_j(t_0^j)\\
\noalign{\vskip3pt}
z_j(t_0^j)\end{matrix}\right),
\qquad
{\bm v_j}=v_j\mkern-3mu\left(\begin{matrix}\cos\theta\\0\\\sin\theta\end{matrix}\right),
\end{equation}
where $y_j(t_0^j)$ and $z_j(t_0^j)$ are random variables, uniformly distributed on the intervals $|y_j(t_0^j)|
\leq w_0\sqrt{|\ln F|}\mkern2mu$ and $|z_j(t_0^j)|\leq \lambda/4$, respectively, and $v_j$ is sampled from the
distribution of atomic speeds [Eq.~(\ref{eqn:speed_dist})]; $\theta$ is the tilt of the atomic beam away
from perpendicular to the cavity axis. The atom moves freely across the cavity after its creation, passing
out of the interaction volume on the plane $x=w_0\sqrt{|\ln F|}$. Thus the interaction volume has a square
rather than circular cross section and measures $2\sqrt{|\ln F|}w_0$ on a side. It is larger than $V_F$
by approximately $30\%$.

Atoms are created in the ground state and returned to the ground state when they leave the interaction volume.
On leaving an atom is disentangled from the system by comparing its probability of excitation with a uniformly
distributed random number $r$, $0\leq r\leq1$, and deciding whether or not it will---anytime in the
future---spontaneously emit; thus, the system state is projected onto the excited state of the leaving atom
(the atom will emit) or its ground state (it will not emit) and propagated forwards in time.

Note that the effects of light forces and radiative heating are neglected. At the thermal velocities
considered, typically the ratio of kinetic energy to recoil energy is of order $10^8$, while the maximum
light shift $\hbar g_{\rm max}$ (assuming one photon in the cavity) is smaller than the kinetic energy by
a factor of $10^7$; even if the axial component of velocity only is considered, these ratios are as high
as $10^4$ and $10^3$ with $\theta\sim10\mkern2mu{\rm mrad}$, as in Figs.~\ref{fig:fig10} and \ref{fig:fig11}.
In fact, the mean intracavity photon number is considerably less than one (Sec.~\ref{sec:photon_number});
thus, for example, the majority of atoms traverse the cavity without making a single spontaneous emission.

Under the atomic beam simulation, the atom number, $N(t)$, and locations ${\bm r_j(t)}$, $j=1,\ldots,N(t)$,
are changing in time; therefore, the atomic state basis is dynamic, growing and shrinking  with $N(t)$. We assume
all atoms couple resonantly to the cavity mode, which is coherently driven on resonance with driving
field amplitude $\cal{E}$. Then, including spontaneous emission and cavity loss, the system is described by
the stochastic master equation in the interaction picture
\begin{eqnarray}
\dot{\rho}&=&{\cal E}[\hat a^{\dag}-\hat a,\rho]+\sum_{j=1}^{N(t)}g({\bm r}_j(t))
[\hat a^{\dag}\hat\sigma_{j-}-\hat a\hat \sigma_{j+},\rho]\nonumber\\
\noalign{\vskip-4pt}
&&+\frac{\gamma}{2}\sum_{j=1}^{N(t)}
\left(2\hat\sigma_{j-}\rho\hat\sigma_{j+}-\hat\sigma_{j+}\hat\sigma_{j-}\rho
-\rho\hat\sigma_{j+}\hat\sigma_{j-}\right)\nonumber\\
\noalign{\vskip6pt}
&&+\kappa\left(2\hat a\rho\hat a^{\dag}-\hat a^{\dag}\hat a\rho
-\rho\hat a^{\dag}\hat a\right),
\label{eqn:master_equation}
\end{eqnarray}
with dipole coupling constants
\begin{equation}
g({\bm r}_j(t))=g_{\rm max}\cos(kz_j(t))\exp\!\left[-\frac{x_j^2(t)+y_j^2(t)}{w_0^2}\right],
\label{eqn:coupling_constant}
\end{equation}
where $\hat a^\dagger$ and $\hat a$ are creation and annihilation operators for the cavity mode, and
$\hat\sigma_{j+}$ and $\hat\sigma_{j-}$, $j=1\ldots N(t)$, are raising and lowering operators for 
two-state atoms.

\subsection{Quantum Trajectory Unraveling}

In principle, the stochastic master equation might be simulated directly, but it is impossible to do
so in practice. Table \ref{tab:parameters} lists effective numbers of atoms $\bar N_{\rm eff}=18$ and
$\bar N_{\rm eff}=13$.  For cut-off parameter $F=0.01$ and an interaction volume of approximately
$1.3\times V_F$ [see the discussion below Eq.~(\ref{eqn:effective_atom_number})], an estimate of the number of interacting
atoms gives $N(t)\sim1.3\times18\bar N_{\rm eff}\approx420$ and $300$, respectively, which means that
even in a two-quanta truncation the size of the atomic state basis ($\sim10^5$ states) is far too large
to work with density matrix elements. We therefore make a quantum trajectory unraveling of
Eq.~(\ref{eqn:master_equation}) \cite{Carmichael93,Dalibard92,Dum92,Gardiner04,Carmichael07b}, where,
given our interest in delayed photon coincidence measurements, conditioning of the evolution upon direct
photoelectron counting records is appropriate: the (unnormalized) conditional state satisfies the nonunitary
Schr\"odinger equation
\begin{equation}
\frac{d|\bar\psi_{\rm REC}\rangle}{dt}=\frac1{i\hbar}\hat H_B(t)|\bar\psi_{\rm REC}\rangle,
\label{eqn:continuous}
\end{equation}
with non-Hermitian Hamiltonian
\begin{eqnarray}
\hat H_B(t)/i\hbar&=&{\cal E}(\hat a^{\dag}-\hat a)+\sum_{j=1}^{N(t)} g({\bm r}_j(t))
(\hat a^{\dag}\hat\sigma_{j-}-\hat a\hat\sigma_{j+})\nonumber \\
&&-\mkern3mu\kappa\hat a^{\dag}\hat a-\frac{\gamma}{2} \sum_{j=1}^{N(t)}\hat\sigma_{j+}\hat\sigma_{j-},
\label{eqn:Hamiltonian1}
\end{eqnarray}
and this continuous evolution is interrupted by quantum jumps that account for photon scattering. There
are $N(t)+1$ scattering channels and correspondingly $N(t)+1$ possible jumps:
\begin{subequations}
\begin{eqnarray}
|\bar\psi_{\rm REC}\rangle\to\hat a|\bar\psi_{\rm REC}\rangle,
\label{eqn:cavity_jump}\\\nonumber
\end{eqnarray}
for forwards scattering---i.e., the transmission of a photon by the cavity---and
\begin{equation}
|\bar\psi_{\rm REC}\rangle\to\hat\sigma_{j-}|\bar\psi_{\rm REC}\rangle,\qquad j=1,\ldots,N(t),
\label{eqn:atom_jump}
\end{equation}
\end{subequations}
for scattering to the side (spontaneous emission). These jumps occur, in time step $\Delta t$, with
probabilities
\begin{subequations}
\begin{equation}
P_{\rm forwards}=2\kappa\langle\hat a^\dag\hat a\rangle_{\rm REC}\Delta t,
\label{eqn:forwards_prob}
\end{equation}
and
\begin{equation}
P_{\rm side}^{(j)}=\gamma\langle\hat\sigma_{j+}\hat\sigma_{j-}\rangle_{\rm REC}\Delta t,\qquad j=1,\ldots,N(t);
\label{eqn:side_prob}
\end{equation}
otherwise, with probability
\end{subequations}
\begin{equation}
1-P_{\rm forwards}-\sum_{j=1}^{N(t)}P_{side}^{(j)},\nonumber
\end{equation}
the evolution under Eq.~(\ref{eqn:continuous}) continues.

For simplicity, and without loss of generality, we assume a negligible loss rate at the cavity input mirror 
compared with that at the output mirror. Under this assumption, backwards scattering quantum jumps need not
be considered. Note that non-Hermitian Hamiltonian (\ref{eqn:Hamiltonian1}) is explicitly time dependent
and stochastic, due to the Monte-Carlo simulation of the atomic beam, and the normalized conditional state is
\begin{equation}
|\psi_{\rm REC}\rangle=\frac{|\bar\psi_{\rm REC}\rangle}{\sqrt{\langle\bar\psi_{\rm REC}|\bar\psi_{\rm REC}\rangle}}.
\end{equation}

\subsection{Two-Quanta Truncation}
Even as a quantum trajectory simulation, a full implementation of our model faces difficulties. The Hilbert space
is enormous if we are to consider a few hundred two-state atoms, and a smaller collective-state basis is inappropriate,
due to spontaneous emission and the coupling of atoms to the cavity mode at unequal strengths. If, on the other hand,
the coherent excitation is sufficiently weak, the Hilbert space may be truncated at the
two-quanta level. The conditional state is expanded as
\begin{widetext}
\begin{equation}
|\psi_{\rm REC}(t)\rangle=|00\rangle+\alpha(t)|10\rangle+\sum_{j=1}^{N(t)}\beta_j(t)|0j\rangle+\eta(t)
|20\rangle+\sum_{j=1}^{N(t)}\zeta_j(t)|1j\rangle+\!\!\sum_{j>k=1}^{N(t)}\vartheta_{jk}(t)|0jk\rangle,
\label{eqn:two_quanta_state}
\end{equation}
\end{widetext}
where the state $|n0\rangle$ has $n=0,1,2$ photons inside the cavity and no atoms excited, $|0j\rangle$ has
no photon inside the cavity and the $j\mkern1mu^{\rm th}$ atom excited, $|1j\rangle$ has one photon inside the
cavity and the $j\mkern1mu^{\rm th}$ atom excited, and $|0jk\rangle$ is the two-quanta state with no photons
inside the cavity and the $j\mkern1mu^{\rm th}$ and $k^{\rm th}$ atoms excited.

The truncation is carried out at the minimum level permitted in a treatment of two-photon correlations. Since
each expansion coefficient need be calculated to dominant order in ${\cal E}/\kappa$ only, the non-Hermitian
Hamiltonian (\ref{eqn:Hamiltonian1}) may be simplified as
\begin{eqnarray}
\hat H_B(t)/i\hbar&=&{\cal E}\hat a^{\dag}+\sum_{j=1}^{N(t)} g({\bm r}_j(t))
(\hat a^{\dag}\hat\sigma_{j-}-\hat a\hat\sigma_{j+})\nonumber \\
&&-\mkern3mu\kappa\hat a^{\dag}\hat a-\frac{\gamma}{2} \sum_{j=1}^{N(t)}\hat\sigma_{j+}\hat\sigma_{j-},
\label{eqn:Hamiltonian2}
\end{eqnarray}
dropping the term $-{\cal E}\hat a$ from the right-hand side. While this self-consistent approximation
is helpful in the analytical calculations reviewed in Sec.~\ref{sec:stationary_atoms}, we do not bother with
it in the numerical simulations.

Truncation at the two-quanta level may be justified by expanding the density operator, along with the master
equation, in powers of ${\cal E}/\kappa$ \cite{Carmichael85,Rice88,Carmichael07c}. One finds that, to
dominant order, the density operator factorizes as a pure state, thus motivating the simplification used
in all previous treatments of photon correlations in many-atom cavity QED \cite{Carmichael91,Brecha99}.
The quantum trajectory formulation provides a clear statement of the physical conditions under which this
approximation holds.

Consider first that there is a fixed number of atoms $N$ and their locations are also fixed. Under weak
excitation, the jump probabilities (\ref{eqn:forwards_prob}) and (\ref{eqn:side_prob}) are very small, and 
quantum jumps are extremely rare. Then, in a time of order $2(\kappa+\gamma/2)^{-1}$, the continuous evolution
(\ref{eqn:continuous}) takes the conditional state to a stationary state, satisfying
\begin{equation}
\hat H_B|\psi_{\rm ss}\rangle=0,
\label{eqn:stationary_state}
\end{equation}
without being interrupted by quantum jumps. In view of the overall rarity of these jumps, to a good
approximation the density operator is
\begin{equation}
\rho_{\rm ss}=|\psi_{\rm ss}\rangle\langle\psi_{\rm ss}|,
\end{equation}
or, if we recognize now the role of the atomic beam, the continuous evolution reaches a quasi-stationary state,
with density operator
\begin{equation}
\rho_{\rm ss}=\overline{\vphantom{\vbox{\vskip8pt}}|\psi_{\rm qs}(t)\rangle\langle\psi_{\rm qs}(t)|\mkern-2mu}
\mkern2mu,
\end{equation}
where $|\psi_{\rm qs}(t)\rangle$ satisfies Eq.~(\ref{eqn:continuous}) (uninterrupted by quantum jumps)
and the overbar indicates an average over the fluctuations of the atomic beam.

This picture of a quasi-stationary pure-state evolution requires the time between quantum jumps to
be much larger than $2(\kappa+\gamma/2)^{-1}$, the time to recover the quasi-stationary state after
a quantum jump has occurred. In terms of photon scattering rates, we require
\begin{equation}
R_{\rm forwards}+R_{\rm side}\ll{\textstyle\frac12}(\kappa+\gamma/2),
\label{eqn:weak_field_limit1}
\end{equation}
where
\begin{subequations}
\begin{eqnarray}
R_{\rm forwards}&=&2\kappa\langle\hat a^\dagger\hat a\rangle_{\rm REC},\label{eqn:forwards_rate}\\
R_{\rm side}&=&\gamma\sum_{j=1}^{N(t)}\langle\hat\sigma_{j+}\hat\sigma_{j-}\rangle_{\rm REC}.
\label{eqn:side_rate}
\end{eqnarray}
\end{subequations}
When considering delayed photon coincidences, after a first forwards-scattered photon is detected, let us say
at time $t_k$, the two-quanta truncation [Eq.~(\ref{eqn:two_quanta_state})] is temporarily reduced by the associated
quantum jump to a one-quanta truncation:
\begin{equation}
|\psi_{\rm REC}(t_k)\rangle\to|\psi_{\rm REC}(t_k^+)\rangle,\nonumber
\end{equation}
where
\begin{equation}
|\psi_{\rm REC}(t_k^+)\rangle=|00\rangle+\alpha(t_k^+)|10\rangle+\sum_{j=1}^{N(t_k)}\beta_j(t_k^+)|0j\rangle,
\label{eqn:one_quanta_state}
\end{equation}
with
\begin{equation}
\alpha(t_k^+)=\frac{\sqrt2\eta(t_k)}{|\alpha(t_k)|},\qquad\beta_j(t_k^+)=\frac{\zeta(t_k)}{|\alpha(t_k)|}.
\end{equation}
%\hfill\break
Then the probability for a subsequent photon detection at $t_k+\tau$ is
\begin{equation}
P_{\rm forwards}=2\kappa|\alpha(t_k+\tau)|^2\Delta t.
\label{eqn:prob_second}
\end{equation}
Clearly, if this probability is to be computed accurately (to dominant order) no more quantum jumps of any
kind should occur before the full two-quanta truncation has been recovered in its quasi-stationary form; in the
experiment a forwards-scattered ``start'' photon should be followed by a ``stop'' photon without any other
scattering events in between. We discuss how well this condition is met by Rempe {\it et al.\/}~\cite{Rempe91}
and Foster {\it et al.\/}~\cite{Foster00a} in Sec.~\ref{sec:photon_number}. Its presumed validity is the
basis for comparing their measurements with formulas derived for the weak-field limit.

\section{Delayed Photon Coincidences for Stationary Atoms}
\label{sec:stationary_atoms}

Before we move on to full quantum trajectory simulations, including the Monte-Carlo simulation of the atomic
beam, we review previous calculations of the delayed photon coincidence rate for forwards scattering with
the atomic motion neglected. Beginning with the original calculation of Carmichael {\it et al.\/}~\cite{Carmichael91},
which assumes a fixed number of atoms, denoted here by $\bar N_{\rm eff}$, all coupled to the cavity mode at strength
$g_{\rm max}$, we then relax the requirement for equal coupling strengths \cite{Rempe91}; finally a Monte-Carlo
average over the spatial configuration of atoms, at fixed density $\varrho$, is taken. The inadequacy
of modeling at this level is shown by comparing the computed correlation functions with the reported data sets. 

\subsection{Ideal Collective Coupling}
For an ensemble of $\bar N_{\rm eff}$ atoms located on the cavity axis and at antinodes of the standing wave,
the non-Hermitian Hamiltonian (\ref{eqn:Hamiltonian2}) is taken over in the form
\begin{eqnarray}
\hat H_B/i\hbar&=&{\cal E}\hat a^\dagger+g_{\rm max}(\hat a^{\dag}\hat J_--\hat a\hat J_+)\nonumber\\
&&-\mkern3mu\kappa\hat a^\dagger\hat a-\frac\gamma4(\hat J_z+N_{\rm eff}),
\end{eqnarray}
where
\begin{equation}
\hat J_{\pm}\equiv\sum_{j=1}^{N_{\rm eff}}\hat\sigma_{j\pm},\qquad\hat J_z\equiv\sum_{j=1}^{N_{\rm eff}}\hat\sigma_{jz}
\end{equation}
are collective atomic operators, and we have written $2\hat\sigma_{j+}\hat\sigma_{j-}=\hat\sigma_{jz}+1$. The conditional state in the
two-quanta truncation is now written more simply as
\begin{widetext}
\begin{equation}
|\psi_{\rm REC}(t)\rangle=|00\rangle+\alpha(t)|10\rangle +\beta(t)|01\rangle+\eta(t)|20\rangle+\zeta(t)
|11\rangle+\vartheta(t)|02\rangle,
\end{equation}
\end{widetext}
where $|nm\rangle$ is the state with $n$ photons in the cavity and $m$ atoms excited, the $m$-atom state being a
collective state. Note that, in principle, side-scattering denies the possibility of using a collective atomic
state basis. While spontaneous emission from a particular atom results in the transition $|n1\rangle\to
\hat\sigma_{j-}|n1\rangle\to|n0\rangle$, which remains within the collective atomic basis, the state $\hat\sigma_{j-}
|n2\rangle$ lies outside it; thus, side-scattering works to degrade the atomic coherence induced by the interaction
with the cavity mode. Nevertheless, its rate is assumed negligible in the weak-field limit [Eq.~(\ref{eqn:weak_field_limit1})],
and therefore a calculation carried out entirely within the collective atomic basis is permitted.

The delayed photon coincidence rate obtained from $|\psi_{\rm REC}(t_k)\rangle=|\psi_{\rm ss}\rangle$ and
Eqs.~(\ref{eqn:one_quanta_state}) and (\ref{eqn:prob_second}) yields the second-order correlation function
\cite{Carmichael91,Brecha99,Carmichael07d}
\begin{widetext}
\begin{equation}
g^{(2)}(\tau)=\left\{ 1-2C_1\frac{\xi}{1+\xi}\frac{2C}{1+2C-2C_1\xi/(1+\xi)}  
\,e^{-\frac{1}{2}(\kappa+\gamma/2)\tau}\! \left[\cos\left(\Omega\tau\right)
\!+\!\frac{\frac{1}{2}(\kappa+\gamma/2)}{\Omega}
\sin\left(\Omega\tau\right)\right]\right\}^2,
\label{eqn:g2_ideal}
\end{equation}
\end{widetext}
with vacuum Rabi frequency
\begin{equation}
\Omega=\sqrt{\bar N_{\rm eff}g_{\rm max}^2-{\textstyle\frac14}(\kappa-\gamma/2)^2},
\label{eqn:vacuum_Rabi_frequency}
\end{equation}
where
\begin{equation}
\xi\equiv2\kappa/\gamma,
\end{equation}
and
\begin{equation}
C\equiv\bar N_{\rm eff}C_1,\qquad C_1\equiv g_{\rm max}^2/\kappa\gamma.
\end{equation}
For $\bar N_{\rm eff}\gg1$, as in Parameter Sets 1 and 2 (Table \ref{tab:parameters}), the deviation from
second-order coherence---i.e., $g^{(2)}(\tau)=1$---is set by $2C_1\xi/(1+\xi)$ and provides a measure of the
single-atom coupling strength. For small time delays the deviation is in the negative direction, signifying
a photon antibunching effect. It should be emphasized that while second-order coherence serves as an unambiguous
indicator of strong coupling in the single-atom sense, vacuum Rabi splitting---the frequency $\Omega$---depends
on the collective coupling strength alone.

\begin{figure}[t]
\hskip-0.2in
\includegraphics[width=3.0in,keepaspectratio=true]{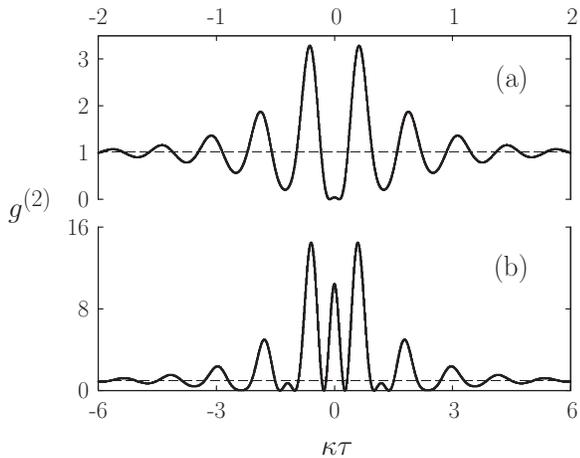}
\caption{Second-order correlation function for ideal coupling [Eq.~(\ref{eqn:g2_ideal})]: (a) Parameter Set 1,
(b) Parameter Set 2.}
\label{fig:fig1}
\end{figure}

Both experiments of interest are firmly within the strong coupling regime, with $2C_1\xi/(1+\xi)=1.2$ for that
of Rempe {\it et al.\/} \cite{Rempe91} ($2C_1=4.6$), and $2C_1\xi/(1+\xi)=4.0$ for that of Foster {\it et al.\/}
\cite{Foster00a} ($2C_1=5.6$). Figure~\ref{fig:fig1} plots the correlation function obtained from Eq.~(\ref{eqn:g2_ideal})
for Parameter Sets 1 and 2. Note that since the expression is a perfect square, the apparent photon bunching of
curve (b) is, in fact, an extrapolation of the antibunching effect of curve (a); the continued nonclassicality
of the correlation function is expressed through the first two side peaks, which, being taller than the central
peak, are classically disallowed \cite{Rice88,Mielke98}. A measurement of the intracavity electric field
perturbation following a photon detection [the square root of Eq.~(\ref{eqn:g2_ideal})] presents a more unified
picture of the development of the quantum fluctuations with increasing $2C_1\xi/(1+\xi)$. Such a measurement may
be accomplished through conditional homodyne detection \cite{Carmichael00,Foster00b,Foster02}.

In Fig.~\ref{fig:fig1} the magnitude of the antibunching effect---the amplitude of the vacuum Rabi
oscillation--- is larger than observed in the experiments by approximately an order of magnitude
(see Fig.~\ref{fig:fig3}). Significant improvement is obtained by taking into account the unequal coupling
strengths of atoms randomly distributed throughout the cavity mode.

\subsection{Fixed Atomic Configuration}
\label{sec:fixed_configuration}
Rempe {\it et al.\/}~\cite{Rempe91} extended the above treatment to the case of unequal coupling strengths, adopting
the non-Hermitian Hamiltonian (\ref{eqn:Hamiltonian2}) while keeping the number of atoms and the atom locations
fixed. For $N$ atoms in a spatial configuration $\{{\bm r}_j\}$, the second-order correlation function takes
the same form as in Eq.~(\ref{eqn:g2_ideal})---still a perfect square---but with a modified amplitude of oscillation
\cite{Rempe91,Carmichael07e}:
\begin{widetext}
\begin{equation}
g^{(2)}_{\{{\bm r}_j\}}(\tau)=\left\{ 1-\frac{[1+\xi(1+C_{\{{\bm r}_j\}})]
S_{\{{\bm r}_j\}}-2C_{\{{\bm r}_j\}}}{1+(1+\xi/2)S_{\{{\bm r}_j\}}}\,
e^{-\frac12(\kappa+\gamma/2)\tau}\!\left[\cos\left(\Omega\tau\right)
+\frac{\frac12(\kappa+\gamma/2)}{\Omega}\sin\left(\Omega\tau\right)\right]\right\}^2,
\label{eqn:g2_fixed_configuration}
\end{equation}
\end{widetext}
with 
\begin{equation}
C_{\{{\bm r}_j\}}\equiv\sum_{j=1}^N C_{1j}, \qquad C_{1j}\equiv g^2({\bm r}_j)/\kappa\gamma,
\end{equation}
\begin{equation}
S_{\{{\bm r}_j\}}\equiv\sum_{j=1}^N \frac{2 C_{1j}}{1+\xi(1+C_{\{{\bm r}_j\}})
-2\xi C_{1j}},
\end{equation}
where the vacuum Rabi frequency is given by Eq.~(\ref{eqn:vacuum_Rabi_frequency}) with effective number of
interacting atoms
\begin{equation}
\bar N_{\rm eff}\to N^{\{{\bm r}_j\}}_{\rm eff}\equiv\sum_{j=1}^Ng^2({\bm r}_j)/g_{\rm max}^2.
\end{equation}

\subsection{Monte-Carlo Average and Comparison with Experimental Results}
\label{sec:Monte_Carlo_average}
In reality the number of atoms and their configuration both fluctuate in time. These fluctuations are readily
taken into account if the typical atomic motion is sufficiently slow; one takes a stationary-atom Monte-Carlo
average over configurations, adopting a finite interaction volume $V_F$ and combining a Poisson average over
the number of atoms $N$ with an average over their uniformly distributed positions ${\bm r}_j$, $j=1,\ldots,N$.
In particular, the effective number of interacting atoms becomes
\begin{equation}
\bar N_{\rm eff}=\overline{N^{\{{\bm r}_j\}}_{\rm eff}},
\end{equation}
where the overbar denotes the Monte-Carlo average. 

Although it is not justified by the velocities listed in Table \ref{tab:parameters}, a stationary-atom
approximation was adopted when modeling the experimental results in Refs.~\cite{Rempe91} and \cite{Foster00a}.
The correlation function was computed as the Monte-Carlo average 
\begin{equation}
g^{(2)}(\tau)=\overline{g^{(2)}_{\{{\bm r}_j\}}(\tau)},
\label{eqn:g2_average1}
\end{equation}
with $g^{(2)}_{\{{\bf r}_j\}}(\tau)$ given by Eq.~(\ref{eqn:g2_fixed_configuration}). In fact, taking a Monte-Carlo
average over {\it normalized\/} correlation functions in this way is not, strictly, correct. In practice, first the
delayed photon coincidence rate is measured, as a separate average, then subsequently normalized by the average photon
counting rate. The more appropriate averaging procedure is therefore
\begin{equation}
g^{(2)}(\tau)=\frac{\overline{\langle\hat a^\dag(0)\hat a^\dag(\tau)
\hat a(\tau)\hat a(0)\rangle_{\{{\bm r}_j\}}}}
{\left(\overline{\langle\hat a^\dag\hat a\rangle_{\{{\bm r}_j\}}}\mkern2mu\right)^2},
\end{equation}
or, in a form revealing more directly the relationship to Eq.~(\ref{eqn:g2_fixed_configuration}), the average
is to be weighted by the square of the photon number:
\begin{equation}
g^{(2)}(\tau)=\frac{\overline{\left(\left\langle \hat a^\dag
\hat a\right\rangle_{\{{\bm r}_j\}}\right)^2g^{(2)}_{\{{\bm r}_j\}}(\tau)}}
{\left(\overline{\langle\hat a^\dag\hat a\rangle_{\{{\bm r}_j\}}}\mkern2mu\right)^2},
\label{eqn:g2_average2}
\end{equation}
where
\begin{equation}
\langle\hat a^\dag\hat a\rangle_{\{{\bm r}_j\}}=\left(\frac{{\cal E}/\kappa}
{1+2C_{\{{\bm r}_j\}}}\right)^2
\label{eqn:photon_number}
\end{equation}
is the intracavity photon number expectation---in stationary state $|\psi_{\rm ss}\rangle$
[Eq.~(\ref{eqn:stationary_state})]---for the configuration of atoms $\{{\bm r}_j\}$.

Note that the statistical independence of forwards-scattering events that are widely separated in time yields
the limit
\begin{equation}
\lim_{\tau\to\infty}g^{(2)}_{\{{\bm r}_j\}}(\tau)\to1,
\end{equation}
which clearly holds for the average (\ref{eqn:g2_average1}) as well. Equation~(\ref{eqn:g2_average2}), on the other
hand, yields
\begin{equation}
\lim_{\tau\to\infty}g^{(2)}(\tau)\to\overline{\left(\left\langle \hat a^\dag
\hat a\right\rangle_{\{{\bm r}_j\}}\right)^2}\bigg{/}\left(\overline{\langle\hat a^\dag\hat a\rangle_{\{{\bm r}_j\}}}
\mkern2mu\right)^2\ge1.
\end{equation}
A value greater than unity arises because while there are fluctuations in $N$ and $\{{\bm r}_j\}$, their correlation
time is infinite under the stationary-atom approximation; the expected decay of the correlation function to unity is
therefore not observed.

\begin{figure}[t]
\hskip-0.2in
\includegraphics[width=3.0in,keepaspectratio=true]{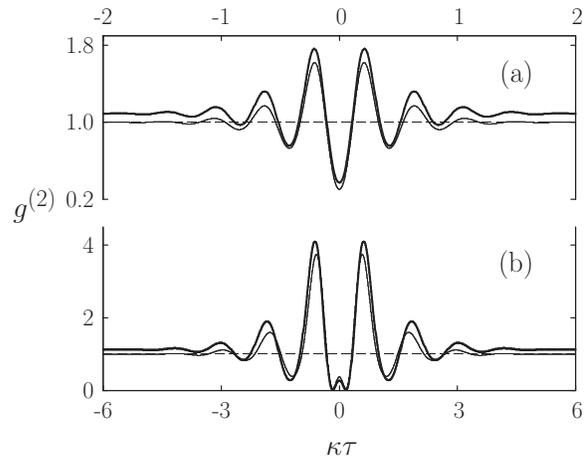}
\caption{Second-order correlation function with Monte-Carlo average over number of atoms $N$ and
configuration $\{{\bm r_j}\}$. The average is taken according to Eq.~(\ref{eqn:g2_average1}) (thin line)
and Eq.~(\ref{eqn:g2_average2}) (thick line) for (a) Parameter Set 1, (b) Parameter Set 2.}
\label{fig:fig2}
\end{figure}

\begin{figure}[b]
\begin{center}
\begin{tabular}{c}
\includegraphics[height=6cm]{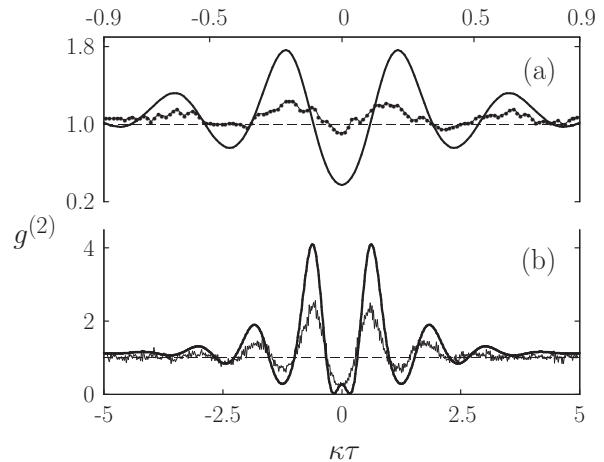}
\end{tabular}
\end{center}
\caption{ \label{fig:fig3}
Second-order correlation function with Monte-Carlo average, Eq.~(\ref{eqn:g2_average2}), over number of atoms $N$
and configuration $\{{\bm r_j}\}$ compared with the experimental data from  (a) Fig.~4(a) of Ref.~\cite{Rempe91}
(Parameter Set 1) and (b) Fig.~4 of Ref.~\cite{Foster00a} (Parameter Set 2).} 
\end{figure}
 
The two averaging schemes are compared in the plots of Fig.~\ref{fig:fig2}, which suggest that atomic beam
fluctuations should have at least a small effect in the experiments; although, just how important they turn
out to be is not captured at all by the figure. The actual disagreement between the model and the data is displayed
in Fig.~\ref{fig:fig3}. The measured photon antibunching effect is significantly smaller than predicted in both
experiments: smaller by a factor of 4 in Fig.~\ref{fig:fig3}(a), as the authors of Ref.~\cite{Rempe91} explicitly
state, and by a factor of a little more than 2 in Fig.~\ref{fig:fig3}(b).

The rest of the paper is devoted to a resolution of this disagreement. It certainly arises from a breakdown
of the stationary-atom approximation as suggested by Rempe {\it et al.\/} \cite{Rempe91}. Physics beyond the
addition of a finite correlation time for fluctuations of $N(t)$ and $\{{\bm r}_j(t)\}$ is needed, however.
We aim to show that the single most important factor is the alignment of the atomic beam.

\section{Delayed Photon Coincidences for an Atomic Beam}
\label{sec:atomic_beam}

We return now to the full atomic beam simulation outlined in Sec.~\ref{sec:cavityQED_atomic_beams}. With the beam
perpendicular to the cavity axis, the rate of change of the dipole coupling constants might be characterized by
the cavity-mode transit time, determined from the mean atomic speed and the cavity-mode waist. Taking the
values of these quantities from Table \ref{tab:parameters}, the experiment of Rempe {\it et al.\/} has
$w_0/\bar v_{\rm source}=182\mkern2mu{\rm nsec}$, which should be compared with a vacuum-Rabi-oscillation decay
time $2(\kappa+\gamma/2)^{-1}=94\mkern2mu{\rm nsec}$, while Foster {\it et al.\/} have $w_0/\bar v_{\rm source}
=66\mkern2mu{\rm nsec}$ and a decay time $2(\kappa+\gamma/2)^{-1}=29\mkern2mu{\rm nsec}$. In both cases, the
ratio between the transit time and decay time is $\sim2$; thus, we might expect the internal state dynamics to
follow the atomic beam fluctuations adiabatically, to a good approximation at least, thus providing a
justifying for the stationary-atom approximation. Figure~\ref{fig:fig3} suggests that this is not so. Our first
task, then, is to see how well in practice the adiabatic following assertion holds.

\subsection{Monte-Carlo Simulation of the Atomic Beam: Effect of Beam Misalignment}
\label{sec:correlation_function}

\begin{figure}[b]
\vskip0.15in
\hskip-0.2in
\includegraphics[width=2.8in,keepaspectratio=true]{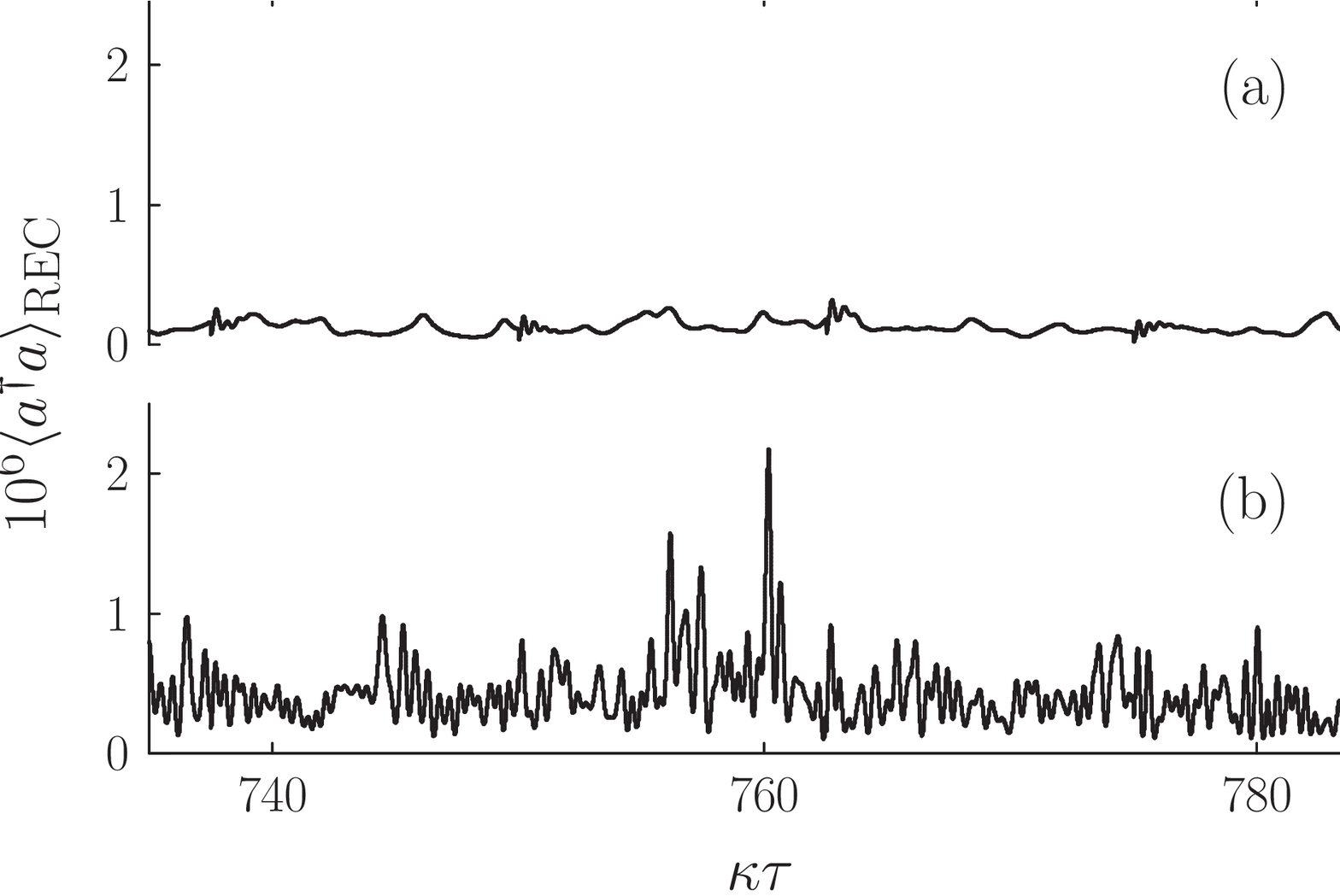}
\caption{Typical trajectory of the intracavity photon number expectation for Parameter Set 1:
(a) atomic beam aligned perpendicular to the cavity axis, (b) with a $9.6\mkern2mu{\rm mrad}$ tilt
of the atomic beam. The driving field strength is ${\mathcal E}/\kappa=2.5\times10^{-2}$.}
\label{fig:fig4}
\vskip0.4in
\hskip-0.2in
\includegraphics[width=2.8in,keepaspectratio=true]{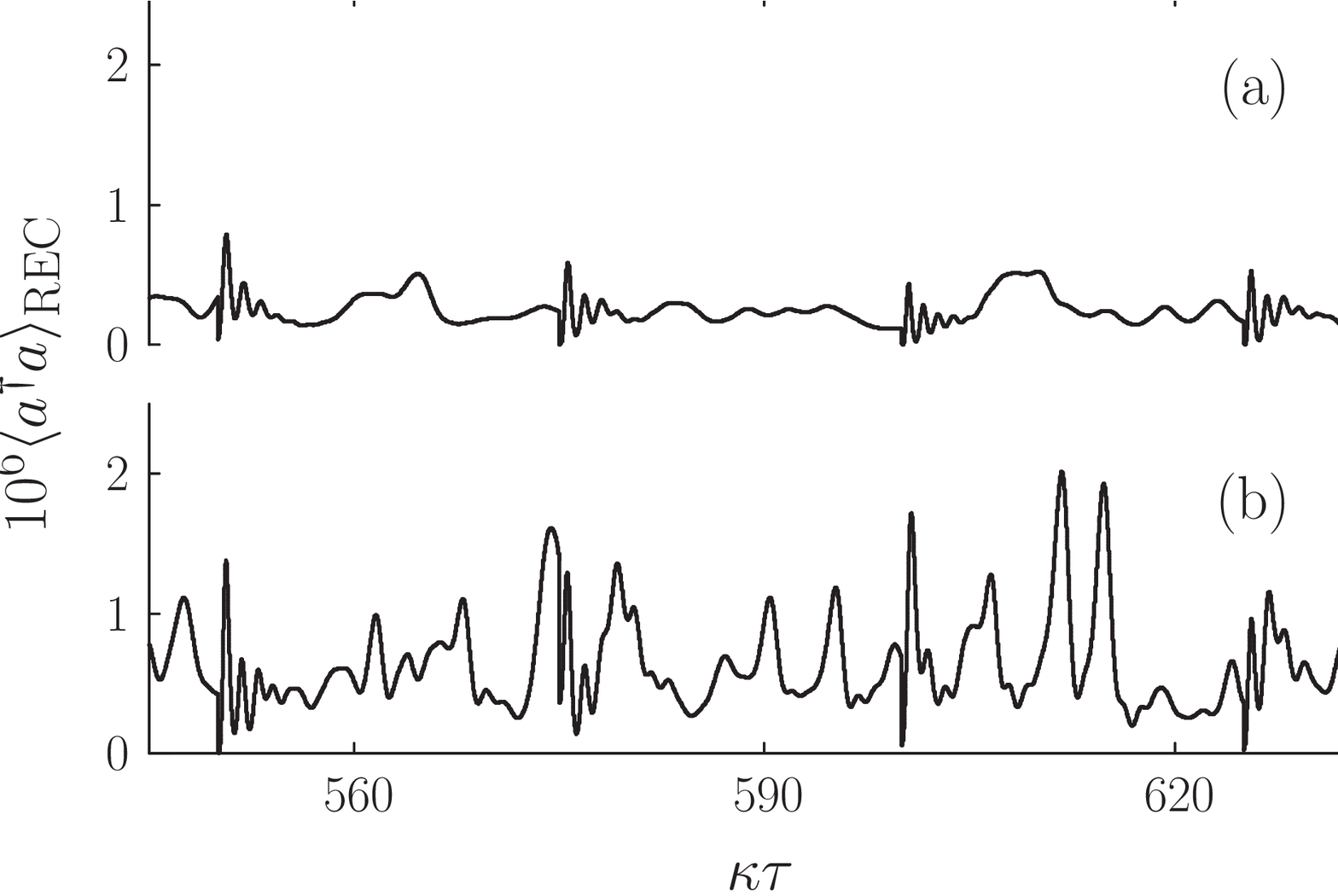}
\caption{As in Fig.~\ref{fig:fig4} but for Parameter Set 2.}
\label{fig:fig5}
\end{figure}

Atomic beam fluctuations induce fluctuations of the intracavity photon number expectation, as illustrated
by the examples in Figs.~\ref{fig:fig4} and \ref{fig:fig5}. Consider the two curves (a) in these figures first,
where the atomic beam is aligned perpendicular to the cavity axis. The ringing at regular intervals along
these curves is the transient response to {\it enforced\/} cavity-mode quantum jumps---jumps
{\it enforced\/} to sample the quantum fluctuations efficiently (see Sec.~\ref{sec:simulation_results}).
Ignoring these perturbations for the present, we see that with the atomic beam aligned perpendicular to the
cavity axis the fluctuations evolve more slowly than the vacuum Rabi oscillation---at a similar rate,
in fact, to the vacuum Rabi oscillation decay. As anticipated, an approximate adiabatic following
is plausible.

\begin{figure}[t]
\hskip-0.2in
\includegraphics[width=2.8in,keepaspectratio=true]{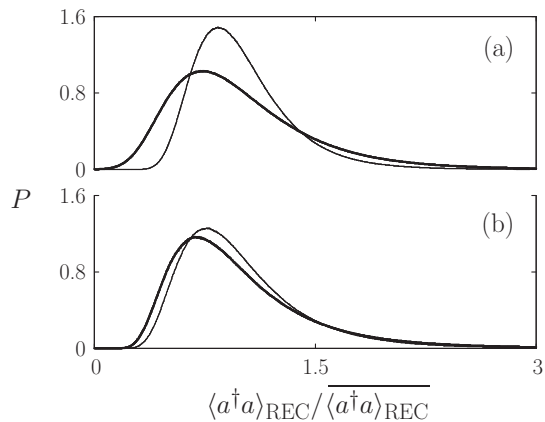}
\caption{Distribution of intracavity photon number expectation with the atom beam perpendicular
to the cavity axis (thin line) and a $9.6\mkern2mu{\rm mrad}$ tilt of the atomic beam (thick line):
(a) Parameter Set 1, (b) Parameter Set 2.}
\label{fig:fig6}
\end{figure}

Consider now the two curves (b); these introduce a $9.6\mkern2mu{\rm mrad}$ misalignment of the atomic beam,
following up on the comment of Foster {\it et al.\/}~\cite{Foster00a} that misalignments as large as~$1^\circ$
($17.45\mkern2mu{\rm mrad}$) might occur. The changes in the fluctuations are dramatic. First, their size
increases, though by less on average than it might appear. The altered distributions of intracavity photon
numbers are shown in Fig.~\ref{fig:fig6}. The means are not so greatly changed, but the variances (measured
relative to the square of the mean) increase by a factor of 2.25 in Fig.~\ref{fig:fig4} and 1.45 in
Fig.~\ref{fig:fig5}. Notably, the distribution is asymmetric, so the most probable photon number lies below
the mean. The asymmetry is accentuated by the tilt, especially for Parameter Set 1 [Fig.~\ref{fig:fig6}(a)].

More important than the change in amplitude of the fluctuations, though, is the increase in their frequency.
Again, the most significant effect occurs for Parameter Set 1 (Fig.~\ref{fig:fig4}), where the frequency with
a $9.6\mkern2mu{\rm mrad}$ tilt approaches that of the vacuum Rabi oscillation itself; clearly, there can
be no adiabatic following under these conditions. Indeed, the net result of the changes from Fig.~\ref{fig:fig4}(a)
to Fig.~\ref{fig:fig4}(b) is that the {\it quantum\/} fluctuations, initiated in the simulation by quantum jumps,
are completely lost in a background of classical noise generated by the atomic beam. It is clear that an atomic
beam misalignment of sufficient size will drastically reduce the photon antibunching effect observed.

For a more quantitative characterization of its effect, we carried out quantum trajectory simulations in a
one-quantum truncation (without quantum jumps) and computed the semiclassical photon number correlation function
\begin{eqnarray}
g^{(2)}_{\rm sc}(\tau)=\frac{\overline{\langle(\hat a^\dag\hat a)(t)\rangle_{\rm REC} 
\langle(\hat a^\dag\hat a)(t+\tau)\rangle_{\rm REC}}}
{\left(\overline{\langle(\hat a^\dag\hat a)(t)\rangle_{\rm REC}}\mkern2mu\right)^2},
\label{eqn:g2_semiclassical}
\end{eqnarray}
where the overbar denotes a time average (in practice an average over an ensemble of sampling times $t_k$). The
photon number expectation was calculated in two ways: first, by assuming that the conditional state
adiabatically follows the fluctuations of the atomic beam, in which case, from Eq.~(\ref{eqn:photon_number}),
we may write
\begin{equation}
\langle(\hat a^\dag\hat a)(t)\rangle_{\rm REC}=\left(\frac{{\cal E/\kappa}}
{1+2C_{\{{\bm r_j}(t)\}}}\right)^2,
\label{eqn:ABC:ad}
\end{equation}
and second, without the adiabatic assumption, in which case the photon number expectation was calculated from
the state vector in the normal way.

\begin{figure}[h]
\hskip-0.2in
\includegraphics[width=3.0in,keepaspectratio=true]{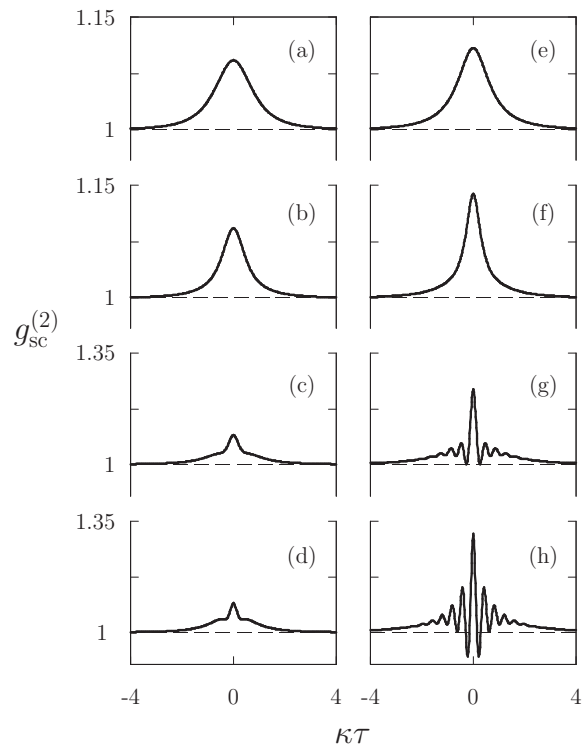}
\caption{Semiclassical correlation function for Parameter Set 1, with adiabatic following of the photon
number (left column) and without adiabatic following (right column); for atomic beam tilts of (a,e) $0\mkern2mu{\rm mrad}$,
(b,f) $4\mkern2mu{\rm mrad}$, (c,g) $9\mkern2mu{\rm mrad}$, (d,h) $13\mkern2mu{\rm mrad}$.}
\label{fig:fig7}
\end{figure}

\begin{figure}[h]
\hskip-0.2in
\includegraphics[width=3.0in,keepaspectratio=true]{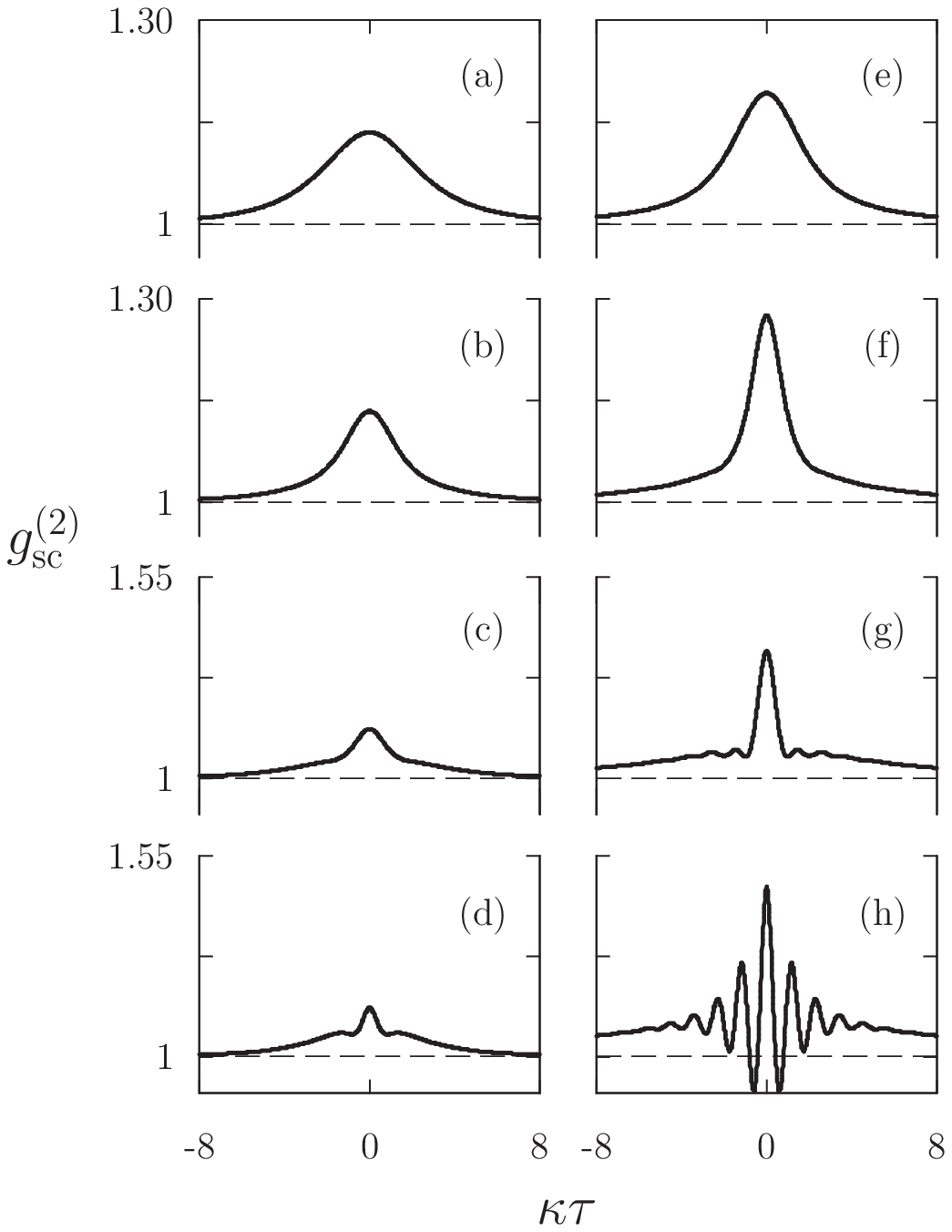}
\caption{As in Fig.~\ref{fig:fig7} but for Parameter Set 2 and atomic beam tilts of (a,e) $0\mkern2mu{\rm mrad}$,
(b,f) $10\mkern2mu{\rm mrad}$, (c,g) $17\mkern2mu{\rm mrad}$, (d,h) $34\mkern2mu{\rm mrad}$.}
\label{fig:fig8}
\end{figure}

Correlation functions computed for different atomic beam tilts according to this scheme are plotted in
Figs.~\ref{fig:fig7} and \ref{fig:fig8}. In each case the curves shown in the left column assume adiabatic
following while those in the right column do not. The upper-most curves [frames (a) and (e)] hold for a beam
aligned perpendicular to the cavity axis and those below [frames (b)--(d) and (f)--(h)] show the effects of
increasing misalignment of the atomic beam. 

A number of comments are in order. Consider first the aligned atomic beam. Correlation times read from the
figures are in approximate agreement with the cavity-mode transit times computed above: the numbers are
$191\mkern2mu{\rm nsec}$ and $167\mkern2mu{\rm nsec}$ from frames (a) and (e), respectively, of
Fig.~\ref{fig:fig7}, compared with $w_0/\bar v_{\rm oven}=182\mkern2mu{\rm nsec}$; and $68\mkern2mu{\rm nsec}$
and $53\mkern2mu{\rm nsec}$ from frames (a) and (e) of Fig.~\ref{fig:fig8}, respectively, compared with
$w_0/\bar v_{\rm oven}=66\mkern2mu{\rm nsec}$. The numbers show a small decrease in the correlation time
when the adiabatic following assumption is lifted (by 10-20\%) but no dramatic change; and there is a
corresponding small increase in the fluctuation amplitude.

Consider now the effect of an atomic beam tilt. Here the changes are significant. They are most evident in frames
(d) and (h) of each figure, but clear already in frames (c) and (g) of Fig.~\ref{fig:fig7}, and frames (b) and (f)
of Fig.~\ref{fig:fig8}, where the tilts are close to the tilt used to generate Figs.~\ref{fig:fig4}(b) and
\ref{fig:fig5}(b) (also to those used for the data fits in Sec.~\ref{sec:simulation_results}). There is first an
increase in the magnitude of the fluctuations---the factors 2.25 and 1.45 noted above---but, more significant, 
a separation of the decay into two pieces: a central component, with short correlation time, and a much broader
component with correlation time larger than $w_0/\bar v_{\rm oven}$. Thus, for a misaligned atomic beam, the
dynamics become notably nonadiabatic.

Our explanation of the nonadiabaticity begins with the observation that any tilt introduces a velocity component along
the standing wave, with transit times through a quarter wavelength of $\lambda/4\bar v_{\rm oven}\sin\theta
=86\mkern2mu{\rm nsec}$ in the Rempe {\it et al.\/} \cite{Rempe91} experiment and $\lambda/4\bar v_{\rm oven}
\sin\theta=60\mkern2mu{\rm nsec}$ in the Foster {\it et al.\/} \cite{Foster00a} experiment. Compared with
the transit time $w_0/\bar v_{\rm oven}$, these numbers have moved closer to the decay times of the vacuum Rabi
oscillation---$94\mkern2mu{\rm nsec}$ and $29\mkern2mu {\rm nsec}$, respectively.  Note that the distances traveled
through the standing wave during the cavity-mode transit, in time $w_0/\bar v_{\rm oven}$, are $w_0\sin\theta=0.53\lambda$
(Parameter Set 1) and $w_0\sin\theta=0.28\lambda$ (Parameter Set 2). It is difficult to explain the detailed shape of
the correlation function under these conditions. Speaking broadly, though, {\it fast atoms\/} produce the central
component, the short correlation time associated with nonadiabatic dynamics, while {\it slow atoms\/} produce the
background component with its long correlation time, which follows from an adiabatic response. Increased tilt brings
greater separation between the responses to fast and slow atoms.

Simple functional fits to the curves in frame (g) of Fig.~\ref{fig:fig7} and frame (f) of Fig.~\ref{fig:fig8}
yield short correlation times of 40-50$\mkern2mu{\rm nsec}$ and $20\mkern2mu{\rm nsec}$, respectively. Consistent
numbers are recovered by adding the decay rate of the vacuum Rabi oscillation to the inverse travel time through
a quarter wavelength; thus, $(1/94+1/86)^{-1}\mkern2mu{\rm nsec}=45\mkern2mu{\rm nsec}$ and $(1/29+1/60)^{-1}
\mkern2mu{\rm nsec}=20\mkern2mu{\rm nsec}$, respectively, in good agreement with the correlation times deduced
from the figures.

The last and possibly most important thing to note is the oscillation in frames (g) and (h) of Fig.~\ref{fig:fig7}
and frame (h) of Fig.~\ref{fig:fig8}. Its frequency is the vacuum Rabi frequency, which shows unambiguously that
the oscillation is caused by a nonadiabatic response of the intracavity photon number to the fluctuations of the
atomic beam. For the tilt used in frame (g) of Fig.~\ref{fig:fig7}, the transit time through a quarter wavelength
is approximately equal to the vacuum-Rabi-oscillation decay time, while it is twice that in  frame (f) of Fig.~\ref{fig:fig8}.
As the tilts used are close to those giving the best data fits in Sec.~\ref{sec:simulation_results}, this would
suggest that atomic beam misalignment places the experiment of Rempe {\it et al.\/}~\cite{Rempe91} further into
the nonadiabatic regime than that of Foster {\it et al.\/}~\cite{Foster00a}, though the tilt is similar in the two
cases. The observation is consistent with the greater contamination by classical noise in Fig.~\ref{fig:fig4}(b) than
in Fig.~\ref{fig:fig5}(b) and with the larger departure of the Rempe {\it et al.\/} data from the stationary-atom
model in Fig.~\ref{fig:fig3}.

\subsection{Simulation Results and Data Fits}
\label{sec:simulation_results}

The correlation functions in the right-hand column of Figs.~\ref{fig:fig7} and \ref{fig:fig8} account for
atomic-beam-induced classical fluctuations of the intracavity photon number. While some exhibit a vacuum Rabi
oscillation, the signals are, of course, photon bunched; a correlation function like that of Fig.~\ref{fig:fig7}(g)
provides evidence of {\it collective\/} strong coupling, but not of strong coupling of the {\it one-atom\/}
kind, for which a photon antibunching effect is needed. We now carry out full quantum trajectory simulations
in a two-quanta truncation to recover the photon antibunching effect---i.e., we bring back the quantum
jumps.

In the weak-field limit the {\it normalized\/} photon correlation function is independent of the amplitude of
the driving field ${\cal E}$ [Eqs.~(\ref{eqn:g2_ideal}) and (\ref{eqn:g2_fixed_configuration})]. The forwards
photon scattering rate itself is proportional to $({\cal E}/\kappa)^2$ [Eq.~(\ref{eqn:photon_number})], and must
be set in the simulations to a value very much smaller than the inverse vacuum-Rabi-oscillation decay time
[Eq.~(\ref{eqn:weak_field_limit1})]. Typical values of the intracavity photon number were $\sim10^{-7}-10^{-6}$.
It is impractical, under these conditions, to wait for the natural occurrence of forwards-scattering quantum jumps.
Instead, cavity-mode quantum jumps are enforced at regular sample times $t_k$ [see Figs.~\ref{fig:fig4}(a) and
\ref{fig:fig5}(a)]. Denoting the record with enforced cavity-mode jumps by $\overline{\vbox{\vskip7.5pt}{\rm REC}
\mkern-2mu}\mkern2mu$, the second-order correlation function is then computed as the ratio of ensemble averages
\begin{equation}
\label{eqn:g2}
g^{(2)}(\tau)=\frac{\overline{\langle(\hat a^\dag\hat a)(t_k)\rangle_{\overline{{\rm REC}\mkern-4mu}}
\mkern4mu
\langle(\hat a^\dag\hat a)(t_k+\tau)\rangle_{\overline{{\rm REC}\mkern-4mu}}\mkern4mu}}
{\left(\overline{\langle(\hat a^\dag\hat a)(t_l)\rangle_{\overline{{\rm REC}\mkern-4mu}}\mkern4mu}
\mkern2mu\right)^{\mkern-2mu 2}}\mkern2mu,
\end{equation}
where the sample times in the denominator, $t_l$, are chosen to avoid the intervals---of duration a few correlation
times---immediately after the jump times $t_k$; this ensures that both ensemble averages are taken in the steady
state. With the cut-off parameter [Eq.~(\ref{eqn:interaction_volume})] set to $F=0.01$, the number of atoms within
the interaction volume typically fluctuates around $N(t)\sim400$-$450$ atoms for Parameter Set 1 and $N(t)\sim280$-$320$
atoms for Parameter Set 2; in a two-quanta truncation, the corresponding numbers of state amplitudes are $\sim90,000$
(Parameter Set 1) and $\sim45,000$ (Parameter Set 2).

\begin{figure}[b]
\hskip-0.2in
\includegraphics[width=3.0in,keepaspectratio=true]{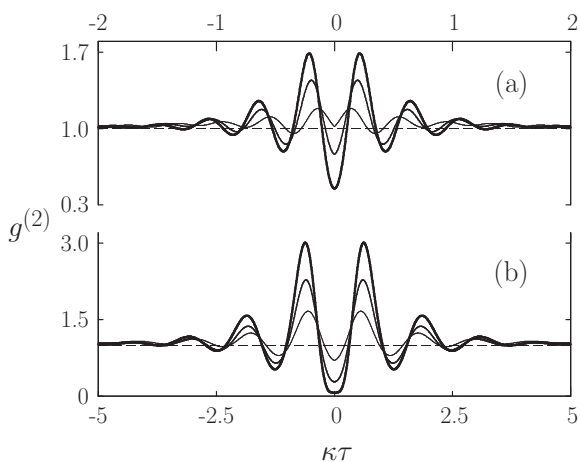}
\caption{Second-order correlation function from full quantum trajectory simulations with a two-quanta truncation:
(a) Parameter Set 1 and $\theta=0\mkern2mu{\rm mrad}$ (thick line), $7\mkern2mu{\rm mrad}$ (medium line),
$12\mkern2mu{\rm mrad}$ (thin line); (b) Parameter Set 2 and $\theta=0\mkern2mu{\rm mrad}$ (thick line),
$10\mkern2mu{\rm mrad}$ (medium line), $17\mkern2mu{\rm mrad}$ (thin line).}
\label{fig:fig9}
\end{figure}

\begin{figure}[t]
\hskip-0.2in
\includegraphics[width=3.0in,keepaspectratio=true]{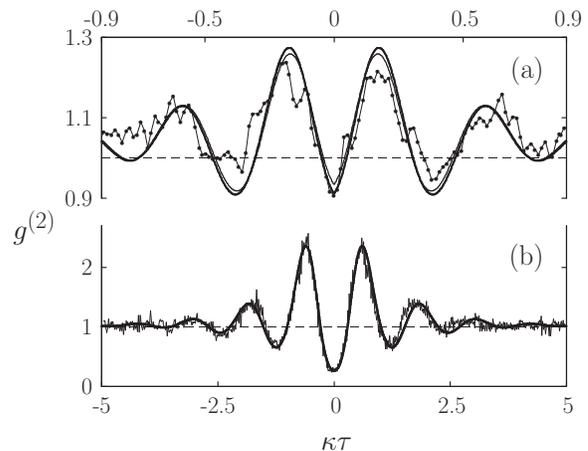}
\caption{Best fits to experimental results: (a) data from Fig.~4(a) of Ref.~\cite{Rempe91} are fitted
with Parameter Set 1 and $\theta=9.7\mkern2mu{\rm mrad}$ (thick line) and $10\mkern2mu{\rm mrad}$ (thin line);
(b) data from Fig.~4 of Ref.~\cite{Foster00a} are fitted with Parameter Set~2 and $\theta=9.55\mkern2mu{\rm mrad}$.
Averages of (a) 200,000 and (b) 50,000 samples were taken with a cavity-mode cut-off $F=0.01$.}
\label{fig:fig10}
\end{figure}

Figure~\ref{fig:fig9} shows the computed correlation functions for various atomic beam tilts. We select from a
series of such results the one that fits the measured correlation function most closely. Optimum tilts are
found to be $9.7\mkern2mu{\rm mrad}$ for the Rempe {\it et al.\/}~\cite{Rempe91} experiment and $9.55\mkern2mu{\rm mrad}$
for the experiment of Foster {\it et al.\/}~\cite{Foster00a}. The best fits are displayed in Fig.~\ref{fig:fig10}.
In the case of the Foster {\it et al.\/} data the fit is extremely good. The only obvious disagreement is that
the fitted frequency of the vacuum Rabi oscillation is possibly a little low. This could be corrected by a small
increase in atomic beam density---the parameter $\bar N_{\rm eff}$---which is only known approximately from the
experiment, in fact by fitting the formula (\ref{eqn:vacuum_Rabi_frequency}) to the data.

The fit to the data of Rempe {\it et al.\/}~\cite{Rempe91} is not quite so good, but still convincing with some
qualifications. Note, in particular, that the tilt used for the fit might be judged a little too large, since the
three central minima in Fig.~\ref{fig:fig10}(a) are almost flat, while the data suggest they should more closely
follow the curve of a damped oscillation. As the thin line in the figure shows, increasing the tilt raises the
central minimum relative to the two on the side; thus, although a better fit around $\kappa\tau=0$ is obtained,
the overall fit becomes worse. This trend results from the sharp maximum in the semiclassical correlation function
of Fig.~\ref{fig:fig7}(g), which becomes more and more prominent as the atomic beam tilt is increased.

The fit of Fig.~\ref{fig:fig10}(b) is extremely good, and, although it is not perfect, the thick
line in Fig.~\ref{fig:fig10}(a), with a $9.7\mkern2mu{\rm mrad}$ tilt, agrees moderately well with the data
once the uncertainty set by shot noise is included, i.e., adding error bars of a few percent (see Fig.~\ref{fig:fig13}).
Thus, leaving aside possible adjustments due to omitted noise sources, such as spontaneous emission---to which
we return in Sec.~\ref{sec:photon_number}---and atomic and cavity detunings, the results of this and the last
section provide strong support for the proposal that the disagreement between theory and experiment presented
in Fig.~\ref{fig:fig3} arises from an atomic beam misalignment of approximately $0.5^\circ$.

\begin{figure}[ht]
\hskip-0.2in
\includegraphics[width=3.0in,keepaspectratio=true]{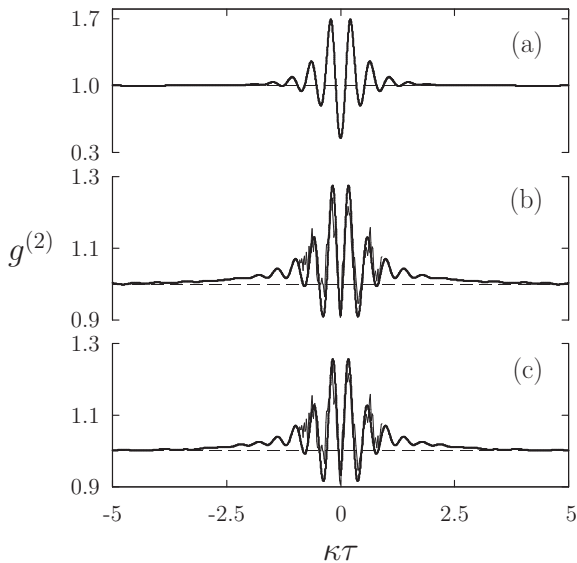}
\caption{Second-order correlation function from full quantum trajectory simulations with a two-quanta basis
for Parameter Set 1 and (a) $\theta=0\mkern2mu{\rm mrad}$, (b) $\theta=9.7\mkern2mu{\rm mrad}$,
(c) $\theta=10\mkern2mu{\rm mrad}$. Averages of (a) 15,000, and (b) and (c) 200,000 samples were taken
with a cavity-mode cut-off $F=0.01$.}
\label{fig:fig11}
\end{figure}

One final observation should be made regarding the fit to the Rempe {\it et al.\/}~\cite{Rempe91} data.
Figure \ref{fig:fig11} replots the comparison made in Fig.~\ref{fig:fig10}(a) for a larger range of time
delays. Frame (a) plots the result of our simulation for a perfectly aligned atomic beam, and frames (b) and
(c) shows the results, plotted in Fig.~\ref{fig:fig10}(a), corresponding to atomic beam tilts of $\theta=9.7
\mkern2mu{\rm mrad}$ and $10\mkern2mu{\rm mrad}$, respectively. The latter two plots are overlayed by the
experimental data. Aside from the reduced amplitude of the vacuum Rabi oscillation, in the presence of
the tilt the correlation function exhibits a broad background arising from atomic beam fluctuations. Notably,
the background is entirely absent when the atomic beam is aligned. The experimental data exhibit just such a
background (Fig.~3(a) of Ref.~\cite{Rempe91}); moreover, an estimate, from Fig.~\ref{fig:fig11}, of the
background correlation time yields approximately $400\mkern2mu{\rm nsec}$, consistent with the experimental
measurement. It is significant that this number is more than twice the transit time, $w_0/\bar v_{\rm oven}
=182\mkern2mu{\rm nsec}$, and therefore not explained by a perpendicular transit across the cavity mode. 
In fact the background mimics the feature noted for larger tilts in Figs.~\ref{fig:fig7} and \ref{fig:fig8};
as mentioned there, it appears to find its origin in the separation of an adiabatic (slowest atoms) from
a nonadiabatic (fastest atoms) response to the density fluctuations of the atomic beam.

Note, however, that a correlation time of $400\mkern2mu{\rm nsec}$ appears to be consistent with a perpendicular
transit across the cavity when the cavity-mode transit time is defined as $2w_0/\bar v_{\rm oven}=364\mkern2mu{\rm nsec}$,
or, using the peak rather than average velocity, as $4w_0/\sqrt\pi\bar v_{\rm oven}=411\mkern2mu{\rm nsec}$; the latter
definition was used to arrive at the $400\mkern2mu{\rm nsec}$ quoted in Ref.~\cite{Rempe91}. There is, of course, some
ambiguity in how a transit time should be defined. We are assuming that the time to replace an ensemble of interacting
atoms with a statistically independent one---which ultimately is what determines the correlation time---is closer
to $w_0/\bar v_{\rm oven}$ than $2w_0/\bar v_{\rm oven}$. In support of the assumption we recall that the
number obtained in this way agrees with the semiclassical correlation function for an aligned atomic beam [Figs.~\ref{fig:fig7}
and~\ref{fig:fig8}, frame (a)].

\subsection{Mean-Doppler-Shift Compensation}
\label{sec:detuning}

Foster {\it et al.\/}~\cite{Foster00a}, in an attempt to account for the disagreement of their measurements
and the stationary-atom model, extended the results of Sec.~\ref{sec:fixed_configuration} to include an
atomic detuning. They then fitted the data using the following procedure: (i) the component of atomic velocity
along the cavity axis is viewed as a Doppler shift from the stationary-atom resonance, (ii) the mean shift is
assumed to be offset by an adjustment of the driving field frequency (tuning to moving atoms) at the time
the data are taken, and (iii) an average over residual detunings---deviations from the mean---is taken in 
the model, i.e., the detuning-dependent generalization of Eq.~(\ref{eqn:g2_fixed_configuration}). The approach
yields a reasonable fit to the data (Fig.~6 of Ref.~\cite{Foster00a}).

The principal difficulty with this approach is that a standing-wave cavity presents an atom with {\it two\/}
Doppler shifts, not one. It seems unlikely, then, that adjusting the driving field frequency to offset one
shift and not the other could compensate for even the average effect of the atomic beam tilt. This difficulty
is absent in a ring cavity, though, so we first assess the performance of the outlined prescription in
the ring-cavity case.

\begin{figure}[t]
\hskip-0.2in
\includegraphics[width=3.0in,keepaspectratio=true]{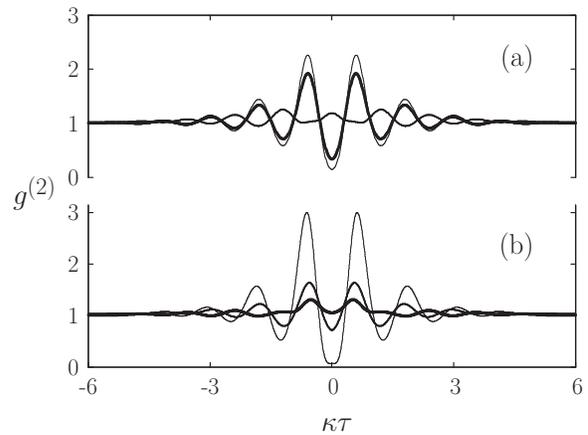}
\caption{Doppler-shift compensation for a misaligned atomic beam in (a) ring and (b) standing-wave cavities
(Parameter Set 2). The second-order correlation function is computed with the atomic beam perpendicular to
the cavity axis (thin line), a $17.3\mkern2mu{\rm mrad}$ tilt of the atomic beam (medium line), and a
$17.3\mkern2mu{\rm mrad}$ tilt plus compensating detuning of the cavity and stationary atom resonances
$\Delta\omega/\kappa=k\bar v_{\rm oven}\sin\theta/\kappa=0.916$ (thick line).}
\label{fig:fig12}
\end{figure}

In a ring cavity, the spatial dependence of the coupling constant [Eq.~(\ref{eqn:coupling_constant})] is
replaced by
\begin{equation}
g({\bm r}_j(t))=\frac{g_{\rm max}}{\sqrt2}\exp(ikz_j(t))\exp\!\left[-\frac{x_j^2(t)+y_j^2(t)}{w_0^2}\right],
\end{equation}
where the factor $\sqrt2$ ensures that the collective coupling strength and vacuum Rabi frequency remain the
same. Figure \ref{fig:fig12}(a) shows the result of a numerical implementation of the proposed mean-Doppler-shift
compensation for an atomic beam tilt of $17.3\mkern2mu{\rm mrad}$, as used in Fig.~6 of Ref.~~\cite{Foster00a}.
It works rather well. The compensated curve (thick line) almost recovers the full photon antibunching effect
that would be seen with an aligned atomic beam (thin line). The degradation that remains is due to the uncompensated dispersion of velocities
(Doppler shifts) in the atomic beam.

For the case of a standing-wave cavity, on the other hand, the outcome is entirely different. This is shown by
Fig.~\ref{fig:fig12}(b). There, offsetting one of the two Doppler shifts only makes the degradation of the photon antibunching effect worse. In
fact, we find that any significant detuning of the driving field from the stationary atom resonance is highly
detrimental to the photon antibunching effect and inconsistent with the Foster {\it et al.\/} data.

\section{Intracavity Photon Number}
\label{sec:photon_number}

The best fits displayed in Fig.~\ref{fig:fig10} were obtained from simulations with a two-quanta truncation and
premised upon the measurements being made in the weak-field limit. The strict requirement of the limit
sets a severe constraint on the intracavity photon number. We consider now whether the requirement is met
in the experiments.

Working from
Eqs.~(\ref{eqn:forwards_rate}) and (\ref{eqn:side_rate}), and the solution to Eq.~(\ref{eqn:stationary_state}),
a fixed configuration $\{{\bm r}_j\}$ of $N$ atoms (Sec.~\ref{sec:fixed_configuration}) yields photon scattering rates
\cite{Carmichael91,Brecha99,Carmichael07c}
\begin{subequations}
\begin{equation}
R_{\rm forwards}=2\kappa\langle\hat a^\dag\hat a\rangle_{\rm REC}=2\kappa\mkern-3mu
\left(\frac{{\cal E/\kappa}}{1+2C_{\{{\bm r_j}\}}}\right)^{\mkern-2mu 2},
\label{eqn:scattering_rate_forwards}
\end{equation}
and
\begin{eqnarray}
R_{\rm side}&=&\gamma\sum_{k=1}^{N}\langle\hat\sigma_{k+}\hat\sigma_{k-}\rangle\nonumber\\
\noalign{\vskip2pt}
&=&\gamma\sum_{k=1}^{N}\left(\frac{g({\bm r}_k)}{\gamma/2}\frac{{\cal E/\kappa}}
{1+2C_{\{{\bm r_j}\}}}\right)^{\mkern-2mu 2}\nonumber\\
\noalign{\vskip2pt}
&=&2C_{\{{\bm r}_j\}}2\kappa\langle\hat a^\dag\hat a\rangle_{\rm REC},
\end{eqnarray}
with ratio
\end{subequations}
\begin{eqnarray}
\frac{R_{\rm side}}{R_{\rm forwards}}=2C_{\{{\bm r}_j\}}=\frac{2N_{\rm eff}^{\{{\bm r}_j\}}
g_{\rm max}^2}{\kappa\gamma}\sim\frac{2\bar N_{\rm eff}g_{\rm max}^2}{\kappa\gamma}.
\label{eqn:scattering_rate_ratio}
\end{eqnarray}
The weak-field limit [Eq.~(\ref{eqn:weak_field_limit1})] requires that the {\it greater\/} of the two rates be much smaller
than $\frac12(\kappa+\gamma/2)$; it is not necessarily sufficient that the forwards scattering rate be low. The side scattering
(spontaneous emission) rate is larger than the forwards scattering rate in both of the experiments being considered---larger
by a large factor of $70$--$80$. Thus, from Eqs.~(\ref{eqn:scattering_rate_forwards}) and (\ref{eqn:scattering_rate_ratio}),
the constraint on intracavity photon number may be written as
\begin{equation}
\langle\hat a^\dag\hat a\rangle\ll\frac{1+\gamma/2\kappa}{8\bar N_{\rm eff}
g^2_{\rm max}/\kappa\gamma},
\label{eqn:weak_field_limit2}
\end{equation}
where, from Table \ref{tab:parameters}, the right-hand side evaluates as $1.2\times10^{-2}$ for Parameter Set 1 and
$4.7\times10^{-3}$ for Parameter Set 2, while the intracavity photon numbers inferred from the experimental count
rates are $3.8\times10^{-2}$ \cite{Rempe91} and $7.6\times10^{-3}$ \cite{Foster00a}. It seems that neither
experiment satisfies condition (\ref{eqn:weak_field_limit2}). As an important final step we should therefore relax
the weak-driving-field assumption (photon number $\sim10^{-7}$--$10^{-6}$ in the simulations) and assess what effect
this has on the data fits; can the simulations fit the inferred intracavity photon numbers as well?

To address this question we extended our simulations to a three-quanta truncation of the Hilbert space with
cavity-mode cut-off changed from $F=0.01$ to $F=0.1$. With the changed cut-off the typical number of atoms in
the interaction volume is halved: $N(t)\sim180$--$220$ atoms for Parameter Set 1 and $N(t)\sim150$--$170$ atoms for
Parameter Set 2, from which the numbers of state amplitudes (including three-quanta states) increase to $1,300,000$
and $700,000$, respectively. The new cut-off introduces a small error in $\bar N_{\rm eff}$, hence in the vacuum Rabi
frequency, but the error is no larger than one or two percent.

At this point an additional approximation must be made. At the excitation levels of the experiments, even
a three-quanta truncation is not entirely adequate. Clumps of three or more side-scattering quantum jumps can
occur, and these are inaccurately described in a three-quanta basis. In an attempt to minimize the error, we
artificially restrict (through a veto) the number of quantum jumps permitted within some prescribed interval
of time. The accepted number was set at two and the time interval to $1\kappa^{-1}$ for Parameter Set~1 and 
$3\kappa^{-1}$ for Parameter Set 2 (the correlation time measured in cavity lifetimes is longer for Parameter Set 2).
With these settings approximately 10\% of the side-scattering jumps were neglected at the highest excitation levels
considered.

\begin{figure}[t]
\vskip0.25cm
\begin{center}
\begin{tabular}{c}
\includegraphics[height=6cm]{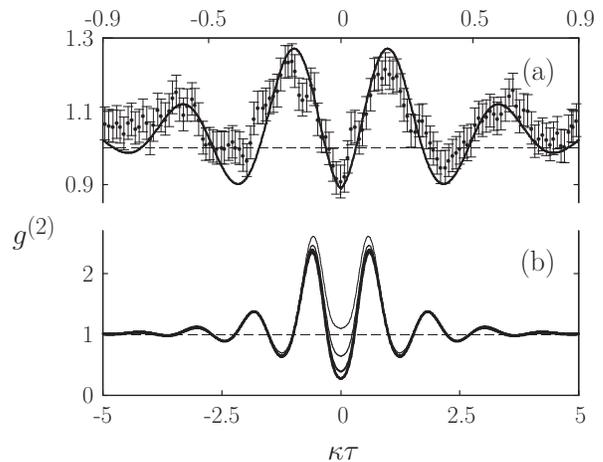}
\end{tabular}
\end{center}
\vskip-0.5cm
\caption{
Second-order correlation function from full quantum trajectory simulations with a three-quanta truncation
and atomic beam tilts as in Fig.~\ref{fig:fig10}: (a) Parameter Set 1, mean intracavity photon number
$\langle a^{\dag} a\rangle=6.7\times 10^{-3}$; (b) Parameter Set 2, mean intracavity photon numbers
$\langle a^{\dag} a\rangle=2.2\times 10^{-4}$, $5.7\times 10^{-4}$, $1.1\times 10^{-3}$, and $1.7\times 10^{-3}$
(thickest curve to thinest curve). Averages of 20,000 samples were taken with a cavity-mode cut-off
$F=0.1$. Shot noise error bars are added to the data taken from Ref.~\cite{Rempe91}. 
\label{fig:fig13}
}
\end{figure}

The results of our three-quanta simulations appear in Fig.~\ref{fig:fig13}; they use the optimal atomic
beam tilts of Fig.~\ref{fig:fig10}. Figure \ref{fig:fig13}(a) compares the simulation with the data of Rempe
{\it et al.}~\cite{Rempe91} at an intracavity photon number that is approximately six times smaller than what
we estimate for the experiment (a more realistic simulation requires a higher level of truncation and is impossible
for us to handle numerically). The overall fit in Fig.~\ref{fig:fig13} is as good as that in Fig.~\ref{fig:fig10},
with a slight improvement in the relative depths of the three central minima. A small systematic disagreement
does remain, however. We suspect that the atomic beam tilt used is actually a little large, while the contribution
to the decoherence of the vacuum Rabi oscillation from spontaneous emission should be somewhat more. We are
satisfied, nevertheless, that the data of Rempe {\it et al.\/}~\cite{Rempe91} are adequately explained by
our model.

Results for the experiment of Foster {\it et al} \cite{Foster00a} lead in a rather different direction. They are
displayed in Fig.~\ref{fig:fig13}(b), where four different intracavity photon numbers are considered. The lowest,
$\langle\hat a^\dagger\hat a\rangle=2.2\times10^{-4}$, reproduces the weak-field result of Fig.~\ref{fig:fig10}(b).
As the photon number is increased, the fit becomes progressively worse. Even at the very low value of $5.7\times10^{-4}$
intracavity photons, spontaneous emission raises the correlation function for zero delay by a noticeable amount. Then
we obtain $g^{(2)}(0)>1$ at the largest photon number considered. Somewhat surprisingly, even this photon number,
$\langle\hat a^\dagger\hat a\rangle=1.7\times10^{-3}$, is smaller than that estimated for the experiment---smaller
by a factor of five. Our simulations therefore disagree significantly with the measurements, despite the near perfect
fit of Fig.~\ref{fig:fig10}(b). The simplest resolution would be for the estimated photon number to be too high.
A reduction by more than an order of magnitude is needed, however, implying an unlikely error, considering
the relatively straightforward method of inference from photon counting rates. This anomaly, for the present,
remains unresolved.

\section{Conclusions}
\label{sec:conclusions}
Spatial variation of the dipole coupling strength has for many years been a particular difficulty for cavity
QED at optical frequencies. The small spatial scale set by the optical wavelength makes any approach to a
resolution a formidable challenge. There has nevertheless been progress made with cooled and trapped atoms
\cite{Hood00,Pinkse00,Boca04,Maunz05,Birnbaum05,Hennrich05}, and in semiconductor systems \cite{Yoshie04,Reithmaier04,Peter05}
where the participating `atoms' are fixed.

The earliest demonstrations of strong coupling at optical frequencies employed standing-wave cavities and thermal
atomic beams, where control over spatial degrees of freedom is limited to the alignment of the atomic beam.
Of particular note are the measurements of photon antibunching in forwards scattering \cite{Rempe91,Mielke98,Foster00a}.
They provide a definitive demonstration of strong coupling at the one-atom level; although many atoms might couple
to the cavity mode at any time, a significant photon antibunching effect occurs only when individual atoms
are strongly coupled.

Spatial effects pose difficulties of a theoretical nature as well. Models that ignore them can point
the direction for experiments, but fail, ultimately, to account for experimental results. In this paper we have
addressed a long-standing disagreement of this kind---disagreement between the theory of photon antibunching
in forwards scattering for stationary atoms in a cavity \cite{Carmichael85,Rice88,Carmichael91,Brecha99,Rempe91}
and  the aforementioned experiments \cite{Rempe91,Mielke98,Foster00a}. {\it Ab initio\/} quantum trajectory
simulations of the experiments have been  carried out, including a Monte-Carlo simulation of the atomic beam.
Importantly, we allow for a misalignment of the atomic beam, since this was recognized as a critical issue in
Ref.~\cite{Foster00a}. We conclude that atomic beam misalignment is, indeed, the most likely reason for the
degradation of the measured photon antibunching effect from predicted results. Working first with a two-quanta
truncation, suitable for the weak-field limit, data sets measured by Rempe {\it et al.\/}~\cite{Rempe91}
and Foster {\it et al.\/}~\cite{Foster00a} were fitted best by atomic beam tilts from perpendicular to the cavity
axis of $9.7\mkern2mu{\rm mrad}$ and $9.55\mkern2mu{\rm mrad}$, respectively.

Atomic motion is recognized as a source of decorrelation omitted from the model used to fit the measurements
in Ref.~\cite{Rempe91}. We found that the mechanism is more complex than suggested there, however. An
atomic beam tilt of sufficient size results in a nonadiabatic response of the intracavity photon number to
the inevitable density fluctuations of the beam. Thus classical noise is written onto the forwards-scattered
photon flux, obscuring the antibunched quantum fluctuations. The parameters of Ref.~\cite{Rempe91}
are particularly unfortunate in this regard, since the nonadiabatic response excites a {\it bunched\/} vacuum
Rabi oscillation, which all but cancels out the antibunched oscillation one aims to measure.

Although both of the experiments modeled operate at relatively low forwards scattering rates, neither is strictly
in the weak-field limit. We have therefore extended our simulations---subject to some numerical constraints---to
assess the effects of spontaneous emission. The fit to the Rempe {\it et al.} data~\cite{Rempe91} was slightly
improved. We noted that the optimum fit might plausibly be obtained by adopting a marginally smaller atomic beam tilt
and allowing for greater decorrelation from spontaneous emission, though a more efficient numerical method
would be required to verify this possibility. The fit to the Foster {\it et al.} data~\cite{Foster00a} was highly
sensitive to spontaneous emission. Even for an intracavity photon number five times smaller than the estimate
for the experiment, a large disagreement with the measurement appeared. No explanation of the anomaly has been
found.

We have shown that cavity QED experiments can call for elaborate and numerically intensive modeling
before a full understanding, at the quantitative level, is reached. Using quantum trajectory methods,
we have significantly increased the scope for realistic modeling of cavity QED with atomic beams.
While we have shown that atomic beam misalignment has significantly degraded the measurements in an
important set of experiments in the field, this observation leads equally to a positive conclusion:
potentially, nonclassical photon correlations in cavity QED can be observed at a level at least ten times
higher than so far achieved.

\section*{Acknowledgements}
This work was supported by the NSF under Grant No.\ PHY-0099576 and by the Marsden Fund of the RSNZ.

%\bibliography{abacqed}

\end{document}